\documentclass[11pt]{article}
\pdfoutput=1 

\usepackage[DIV13]{typearea}
\usepackage{amsmath}
\usepackage{amsfonts}
\usepackage{amssymb}
\usepackage[titletoc]{appendix}
\usepackage{array}
\usepackage{bbold} 
\usepackage{bbm}
\usepackage{booktabs}
 \usepackage{cancel}
 \usepackage{cite}
\usepackage[usenames,dvipsnames]{color}
\usepackage{epsfig}
\usepackage{fancyhdr}
\usepackage{feynmp}
\usepackage[T1]{fontenc}
\usepackage{framed}
\usepackage[left=.9in, right=.9in]{geometry}
\usepackage{graphicx}
\usepackage{hhline} 
\RequirePackage[colorlinks=true,urlcolor=blue,anchorcolor=blue,citecolor=blue,filecolor=blue,
               linkcolor=blue,menucolor=blue,linktocpage=true,pdfproducer=medialab]{hyperref}
\usepackage[utf8]{inputenc}
\usepackage{mathrsfs}
\usepackage{mathtools}
\usepackage{multicol}
\usepackage{multirow}
\usepackage{indentfirst}
\usepackage{longtable}
 \usepackage{lscape}
\usepackage{pbox}
\usepackage{pdfpages}
\usepackage{subcaption}
\usepackage{rotating}
\usepackage{slashed}
\usepackage[DIV13]{typearea}
\usepackage[normalem]{ulem}
\usepackage{units}
\usepackage{enumitem}
\usepackage{soul}
\usepackage{placeins}

\usepackage{catchfile}
\newcommand{\getenv}[2][]{%
  \CatchFileEdef{\temp}{"|kpsewhich --var-value #2"}{}%
  \if\relax\detokenize{#1}\relax\temp\else\let#1\temp\fi
}
\getenv[\USER]{USER}

\allowdisplaybreaks
\newcommand{\g}{\gamma}
\newcommand{\m}{\mu}

\newcommand{\s}{\sigma}

\renewcommand{\m}{\mu}
\newcommand{\n}{\nu}

\renewcommand{\L}{\mathcal{L}}

\renewcommand{\to}{\rightarrow}

\newcommand{\hc}{\mathrm{h.c.}}

\newcommand{\beq}{\begin{equation}}
\newcommand{\eeq}{\end{equation}}
\newcommand{\bea}{\begin{eqnarray}}
\newcommand{\eea}{\end{eqnarray}}
\renewcommand{\[}{\begin{equation}}
\renewcommand{\]}{\end{equation}}
\definecolor{orange}{rgb}{1,0.5,0}

\newcommand{\blue}[1]{\color{blue} #1 \color{black}}

\newcommand{\Ll}{\Lambda}

\newcommand{\psibar}{\bar{\psi}}
\newcommand{\Psibar}{\bar{\Psi}}
\newcommand{\qbar}{\bar{q}}

\newcommand{\Tbar}{\bar{T}}
\newcommand{\Bbar}{\bar{B}}

\newcounter{diagram}

\newcommand{\email}[1]{\href{mailto:#1}{\tt #1}}

\renewcommand{\to}{\rightarrow}
\newcommand{\LRdel}{\Delta}
\newcommand{\spurion}{\Gamma}

\def\Ds{\slashed{D}}

\def\Lam{\Lambda}
\newcommand{\paren}[1]{\left( #1 \right)}

\newcommand{\nn}{\nonumber}
\renewcommand{\to}{\longrightarrow}

\renewcommand{\[}{\begin{equation}}
\renewcommand{\]}{\end{equation}}
\newcommand{\bmat}{\begin{pmatrix}}
\newcommand{\emat}{\end{pmatrix}}

\newcolumntype{C}[1]{>{\centering\let\newline\\\arraybackslash\hspace{0pt}}m{#1}}

\begin{document}
\vspace*{-1cm}
\phantom{hep-ph/***} 
{\flushleft
{\blue{IFT-UAM/CSIC-17-103}}
\hfill{\blue{FTUAM-17-24}}
}
\vskip 1.5cm
\begin{center}
{\LARGE  The Axion and the Goldstone Higgs\\[3mm] }
\vskip .3cm
\end{center}
\vskip 0.5  cm
\begin{center}
{\large I.~Brivio}~$^{a)}$,
{\large M.B.~Gavela}~$^{b)}$,
{\large S.~Pascoli}~$^{c)}$, 
{\large R.~del Rey}~$^{b)}$,
{\large S.~Saa}~$^{b)}$
\\
\vskip .7cm
{\footnotesize
$^{a)}$~
Niels Bohr International Academy and Discovery Center, Niels Bohr Institute, University of Copenhagen, DK-2100 Copenhagen, Denmark\\
\vskip .1cm
$^{b)}$~
Departamento de F\'isica Te\'orica and Instituto de F\'{\i}sica Te\'orica, IFT-UAM/CSIC,\\
Universidad Aut\'onoma de Madrid, Cantoblanco, 28049, Madrid, Spain\\
\vskip .1cm
$^{c)}$~
Institute for Particle Physics Phenomenology, Department of Physics, Durham University,
South Road, Durham DH1 3LE, U.K.

\vskip .5cm
\begin{minipage}[l]{.9\textwidth}
\begin{center} 
\textit{E-mail:} 
\email{ilaria.brivio@nbi.ku.dk},
\email{belen.gavela@uam.es},
\email{silvia.pascoli@durham.ac.uk},
\email{rocio.rey@uam.es},
\email{sara.saa@uam.es}
\end{center}
\end{minipage}
}
\end{center}

\begin{abstract}
We  consider  the  renormalizable $SO(5)/SO(4)$ $\sigma$-model,  in which the Higgs particle has a pseudo-Nambu-Goldstone boson character, and    explore what the minimal field extension required to implement the Peccei-Quinn symmetry (PQ) is, within the partial compositeness scenario. It turns out that the minimal model does not require the enlargement of the exotic fermionic sector, but only the addition of a singlet scalar:  it is sufficient that the exotic fermions involved in partial compositeness and the singlet scalar become charged under Peccei-Quinn transformations.
We explore the phenomenological predictions for photonic signals in axion searches for all models discussed.
Because of the constraints imposed on the  exotic fermion sector by the Standard Model fermion masses, the expected range of allowed axion-photon couplings turns out to be generically narrowed with respect to that of standard invisible axion models, impacting the experimental quest. 
 
\end{abstract}

\vskip 1cm

\pagebreak
\tableofcontents

\pagebreak

%
%
\section{Introduction}
\label{Sect:Intro}

The Standard Model of Particle Physics (SM) describes amazingly well the interactions among the known particles. Describing does not mean
understanding, though. One crucial issue not understood is the SM fundamental state, that is, its vacuum. The
vacuum of the SM strong interactions is characterized by  a parameter $\theta$ whose physical value needs to
be strongly adjusted for no apparent reason, $\theta\le 10^{-10}$; it is the so-called ``strong CP
problem'', a major long-standing fundamental SM puzzle.  
Furthermore, the solutions proposed to another fine-tuning problem of the SM, the so-called electroweak hierarchy problem, often contribute to the strong CP problem: most  beyond the SM theories  devised to solve the former induce unacceptably large electroweak quantum corrections to $\theta$.

The standard dynamical solution to the strong CP problem of QCD is  based on extending the SM  with a spontaneously broken global axial ``Peccei-Quinn'' (PQ) symmetry, $U(1)_{\mathrm{PQ}}$, whose associated (pseudo)Nambu-Goldstone boson (pGB) is the ``axion'' $a(x)$~\cite{Peccei:1977hh, Weinberg:1977ma, Wilczek:1977pj}.   In its most economic and traditional realization, which is the one to be considered here, the matter sector of the SM needs to be extended, but not its gauge group.\footnote{A second main avenue not considered here is that in which the strong sector of the SM  gauge group is enlarged by a new gauge sector with scales which are typically much higher than the QCD scale. This may allow for relatively low values of the new physics scale $f_a${~\cite{Rubakov:1997vp, Berezhiani:2000gh, Hook:2014cda, Gherghetta:2016fhp}} and has been rarely explored.}  As a consequence,  independently of the model details the product of the axion mass $m_a$  and scale $f_a$ obeys then the relation
\beq
\label{mafa}
m_a^2\,f_a^2\,\approx\,m_\pi^2\,f_\pi^2\,\frac{m_um_d}{(m_u+m_d)^2},
\eeq
where $\pi$ denotes the pion. The right-hand side of this equation is of the order of the QCD scale, whose non-perturbative effects break the global PQ symmetry and are responsible for the non-vanishing axion mass.  The 
downside is that, as the axion coupling to ordinary matter is 
proportional to $1/f_a$, phenomenological and astrophysical constraints~\cite{Ayala:2014pea, Vinyoles:2015aba} tend to require then extreme values for $f_a$, typically : $10^9< f_a <10^{14}$ GeV, that is, $10^{-7}<m_a<10^{-2}$ eV.  These models are known by the general name of ``invisible axion'' theories.  This axion may also be an excellent candidate   ``dark matter'' particle. The two paradigmatic examples of invisible axion theories require to add to the SM spectrum a scalar singlet $S$ whose vacuum expectation value (vev) breaks PQ and sets the scale $f_a$: the DFSZ model~\cite{Zhitnitsky:1980tq, Dine:1981rt}, where a second Higgs doublet is also added, and the KSVZ model~\cite{Kim:1979if, Shifman:1979if}  which requires extra vectorial fermions instead. They guide the current very intense experimental search. Note that the ADMX experiment~\cite{Asztalos:2009yp, Carosi:2013rla, DePanfilis:1987dk, Wuensch:1989sa, Hagmann:1990tj, Asztalos:2003px} has started to enter the critical region of axion-photon favoured by the invisible axion; this has tantalized the particle physics community, as a discovery would be revolutionary. Axion-like photonic couplings 
approaching the invisible axion window are also being searched for by CAST~\cite{Anastassopoulos:2017ftl}, and the International Axion Observatory
IAXO is in fast preparation~\cite{Irastorza:2011gs, Carosi:2013rla}, together with other laboratory experiments such as ALPS-II~\cite{Bahre:2013ywa}.

{\it The major drawback of extensions of the Standard Model which embed an invisible axion is that they are strongly fine-tuned in their scalar potential, as $f_a$ is in general many orders of magnitude larger than the observed electroweak scale. }
Indeed, the Higgs and axion sectors are not watertight but communicate through the scalar potential, which includes $S$-Higgs interactions which would pull the Higgs mass towards the high scale. The range of $f_a$ mentioned above may be loosened in invisible axion models by assuming several exotic matter representations with ad-hoc cancellations of their contributions to the axion-photon-photon couplings~\cite{DiLuzio:2016sbl,  DiLuzio:2017pfr}, avoiding then some of the most stringent astrophysical constraints.
Nevertheless,  purely hadronic bounds still hold even in this case, which still imply a few orders of magnitude difference between the value of $f_a$ and the electroweak scale~\cite{Fukuda:2015ana}. A coherent picture of the solution to the strong CP problem is thus missing.~\footnote{In addition, quantum gravitational effects violate global symmetries such as the PQ symmetry, and Planck-suppressed higher dimension operators are a threat unless the $f_a$ scale is small. This aspect of invisible axion models with very large $f_a$ is not addressed here.}

It is obviously possible to implement the (high-scale) invisible axion solution to the strong CP problem, without the Higgs mass suffering from the electroweak hierarchy problem,  though, assuming that the Higgs mass is protected by some symmetry: supersymmetric models at the electroweak scale and the so-called ``composite Higgs models" are examples of it. The latter are within the class of models in which the Higgs particle   is protected from putative higher scales  via  a pseudo-Nambu-Goldstone boson ancestry~~\cite{Kaplan:1983fs, Georgi:1984af, Dugan:1984hq}, and we focus in this paper on  this avenue (named in what follows ``Goldstone Higgs'' for brevity).   
In their most economical realization, the gauge group is just the SM one while both  the Higgs  and the longitudinal degrees of freedom of the electroweak bosons originate as the GBs  from  a global $SO(5)$ symmetry~\cite{Agashe:2004rs, Contino:2006qr} spontaneously broken to $SO(4)$ at some high scale $\Lambda_s$.   

It would be straightforward and trivial to extend such  theories
so as to implement on them a PQ solution, by simply adding at a higher scale supplementary matter fields specifically for that purpose. Here, instead, we focus on the minimal possible extension.  That is, we explore within $SO(5)/SO(4)$ whether the minimal exotic fermionic setup of partial compositeness suffices to the purpose. 
Indeed, a  recurrent  characteristic in Goldstone Higgs models is the implementation of partial compositeness via exotic  fermion representations which are vectorial with respect to the SM gauge group $SU(2)_L$, in the sense that the left and right-handed components are in equal representations of $SU(2)_L$, much as in KSVZ invisible axion theories. We will take advantage of this fact in this work, and it will be shown that 
a minimal scalar extension is enough to make the models PQ invariant. 
 
In partial compositeness, the global symmetry and spectra forbid tree-level Yukawa couplings and the SM masses are mediated instead by the exotic vectorial fermions. This imposes stringent relations among the  parameters and couplings of that exotic fermion sector, which will be shown to point to a reduced phenomenological parameter space when a Peccei-Quinn symmetry is implemented using those same fermions.
 
We will formulate first the question using a complete renormalizable model~\cite{Feruglio:2016zvt}, which in its scalar part is a linear sigma
model including a new scalar particle, $\sigma$, singlet under the gauge group: the linear $\sigma$ model for the composite Higgs. The model can be considered either as an ultimate theory describing
elementary fields (instead of composite ones), or as a renormalizable version of a deeper dynamics, much as the
linear $\sigma$ model is to QCD; in the limit of very heavy $\sigma$ mass, the non-linear regime is reached.  
A clear advantage of using first a complete renormalizable model for a Higgs with GB ancestry is that it allows to gauge how costly the implementation of the PQ symmetry for composite Higgs constructions is, in terms of extending its spectrum and in particular its scalar sector, a task not feasible or at least very obscure in non-renormalizable  formulations. Moreover, the need in invisible axion constructions to strongly raise $f_a$ above the electroweak scale suggests its pairing with the limit of a very heavy $\sigma$ particle, as the mass of the latter is not protected and a light $\sigma$ could raise issues of fine-tuning.
At this point, an overview of the scales involved is pertinent:
\begin{itemize}
\item The electroweak (EW) scale $v$.
\item The  Goldstone-boson scale $f$ associated to the physical Higgs $h$, whose value does not need to coincide with $v$, and is typically expected to be in  the TeV range, $f\gtrsim 500$ GeV~\cite{Agashe:2004rs, Contino:2006qr}. It is the analogous of $f_\pi$ in QCD.
\item The scale $\Lambda_s$ of the high-energy strong dynamics responsible for the pseudo-Goldstone boson nature of the Higgs field, with $\Lambda_s\leq 4\pi f$~\cite{Manohar:1983md}, and in consequence approximately in the $1- 10^2$ TeV range. This is the overall scale of the Higgs theory  and, as such, this sets intuitively the scale  of masses expected for the exotic fermions, much as in QCD the overall scale of the theory corresponds to the proton mass. 
\item The  axion scale $f_a$, which is many orders of magnitude larger than any of the above, given the experimental and observational constraints subject to Eq.~(\ref{mafa}). Such a large value is naturally accommodated when it corresponds to the vev of a scalar $S$ which is a singlet of both the SM and $SO(5)$.
\item The $\sigma$ mass, which can range from few hundred of GeV to infinitely heavy in the strong interacting regime. The latter avoids fine-tuning issues by construction. Because the sigma particle is not a goldstone mode, it can be made heavy without affecting the GB scale $f$ nor the Higgs mass, which are controlled by $SO(5)$-breaking effects (see Ref~\cite{Feruglio:2016zvt} for the Lagrangian). This is analogous to the nonlinear limit of the sigma model in QCD. This  mass scale is absent in non-linear realizations, which are akin to a very heavy $\sigma$ decoupled from the spectrum, much as the chiral Lagrangian for QCD with a light pion decay constant $f_\pi$ corresponds to the infinite mass limit of the renormalizable linear sigma model for QCD.  
Alternatively, when the $\sigma$ particle is present in the spectrum, the scalar potential may tend to homogenize the size of all singlet parameters, e.g. the $\sigma$ mass and $f_a$. 

Note that when minimally extending the renormalizable model to encompass an axion solution to the strong CP problem, the scalar $\sigma$ cannot be charged under the PQ symmetry as it belongs to a real scalar five-plet of $SO(5)$; $S$ and $\sigma$ are thus independent fields.

\end{itemize}

By construction, \emph{the mass of  the Goldstone  Higgs will not be destabilized by the high scale $f_a$ as far as the model preserves the approximate $SO(5)/SO(4)$ symmetry pattern}. 
Nevertheless, the   simultaneous presence of the very high scale $f_a$ and the lower scales immediately raises the question of whether  some of the axion-related parameters of the Lagrangian may have to be fine-tuned, e.g. in the scalar potential.  In particular, 
the exotic vectorial fermion masses in  traditional invisible axion models ``\`a la KSVZ'' stem from the vev of $S$, $\langle S \rangle \sim f_a$, a fact  that could be  in tension with the requirement of  
much lighter fermionic states in composite Higgs models. This issue will be addressed discussing the technical naturalness of (dimensionless) mass parameters in the exotic fermion sector.
 
A precision is pertinent on the size of the exotic fermion masses, though. The comment above expecting them to be of order $\Lambda_s$ --that is, not very far from the TeV region-- is the natural expectation if no higher scales were present in Nature to which the system is sensitive. Here instead, in particular with a renormalizable model which is in itself ultraviolet complete, one cannot exclude that some $SO(5)$-invariant mass parameters may be of the order of the highest scale in the theory, the $f_a$ scale, as quantum corrections may  equalize all singlet scales, depending in particular on the couplings in the scalar potential. For instance, fermionic masses of $\mathcal{O}(f_a)$ could be possible for singlet fermions. In fact these contributions are not expected to destabilize the relative size of the Higgs mass, as the latter must be proportional to symmetry-breaking parameters. However, it remains to be verified, with a complete one-loop study of the scalar potential, whether the overall scale $f$ would be pushed to large values in the presence of very heavy fermion singlets. In Ref.~\cite{Pomarol:2012qf} it was shown that  some of the exotic fermion masses could indeed be larger than the overall TeV scale and still comply with the Higgs and fermion masses, as far as other fermions were lighter than that scale. For the sake of completeness, such particular cases will be included in the discussion below, although in most of our study  we will take the most conservative option of assuming implicitly ``light'' exotic fermion masses $\sim\Lambda_s$, unless otherwise stated.  In any case, the phenomenological predictions for axion-photon couplings will be independent of those fermion mass values. 

One last comment is in order with respect to the relevant scales in the theory, involving now the hierarchy between the PQ scale and possible new physics arising, for instance, at the Planck scale. It has often been argued indeed that all global symmetries may be violated by non-perturbative quantum gravitational effects (associated for instance to black holes or to wormhole physics),  and these gravitational corrections dangerous for axion models in which the PQ scale is not far from the Planck scale. These effects are typically formulated in terms of effective operators, suppressed by the Planck scale, that describe the possible new physics contributions to the axion potential that would shift the vacuum and leave the strong CP problem unsolved.   Nevertheless, the idea that gravity breaks all global symmetries is indeed an assumption and sometimes an incorrect one - at least at the Lagrangian level.~\footnote{For example, orbifold compactifications of the heterotic string have discrete symmetries that prevent the presence of some higher dimension operators, and this can strongly and safely suppress the dangerous effects under discussion~\cite{Butter:2005wr}.} For example, very recently the impact of wormholes  has been clarified and quantified in Ref.~\cite{Alonso:2017avz}. These non-perturbative effects turn out to be extremely suppressed by an exponential dependence on the gravitational instanton action, and they are harmless even with high axion scales. The demonstration relied only on assuming that the spontaneous breaking of the PQ symmetry is implemented through the vev of a scalar field, and thus it directly applies to our model. Nevertheless, if the issue were to appear anyway, a number of interesting proposals exist where PQ symmetry arises automatically as a consequence of gauge invariance in an extended gauge setup. While many of these proposals are very different to the model here presented (as they are composite axion or SUSY setups), others rely on a KSVZ-type axion construction~\cite{DiLuzio:2017tjx} and may possibly be made compatible with the model presented in this paper.

For concreteness, we will explore first the renormalizable model of Ref.~\cite{Feruglio:2016zvt}, examining the minimal matter extensions that ensure PQ symmetry.   We remind here that the question of the stability of the Higgs mass in such a setup has been already clarified in the literature~\cite{Feruglio:2016zvt, Merlo:2017sun}. The scalar potential in that model for the Higgs ($h$) and $\sigma$ particles can be written as
\begin{equation}
V(h,\sigma) = \lambda\left(h^2+\sigma^2-f^2\right)^2+\alpha f^3\,\sigma-\beta f^2\,h^2\,,
 \label{Laghs}\,
\end{equation}
where $\lambda$ is the coefficient of the dominant SO(5)-invariant term, while $\alpha$ and $\beta$ are the coefficients of the counterterms that reabsorb relevant divergences. The presence of $\alpha$ and $\beta$ breaks the global  $SO(5)$ symmetry: the smallness of these parameters is therefore ensured by 't Hooft's naturalness principle and guarantees that the symmetry remains approximate. 

A consistent electroweak (EW) symmetry breaking requires both scalars $h$ and $\sigma$ to acquire a non-vanishing vev, 
respectively dubbed as $v$ and $v_\sigma$ below.  
Note that the  vev of $h$ is identified with the electroweak scale since it can be related to the Fermi constant precisely as in the SM. For $\alpha, \beta \neq 0$ and assuming $v\ne0$,  it results 
\begin{equation}
v_{\sigma}^2=f^2\frac{\alpha^2}{4\beta^2} \quad , \qquad v^2 = f^2\left(1-\frac{\alpha^2}{4\beta^2}+\frac{\beta}{2\lambda}\right)\,, 
\label{vevs}
\end{equation}
satisfying the condition 
\begin{equation}
v^2+v_{\sigma}^2=f^2\left(1+\beta/2\lambda\right)\,, 
\label{vevs2}
\end{equation}
which indicates that the $SO(5)$ vev is ``renormalized'' by the
$\beta$ term in the potential. This leads to the following expressions for the scalar mass eigenvalues:
    \begin{equation} 
m^2_\text{heavy, light}=
	4\lambda f^2\left\{ \left(1+\frac{3}{4}\frac{\beta}{\lambda} \right) \pm 
	\left[1+ \frac{\beta}{2 \lambda}\left(1 +\frac{\alpha^2}{2 \beta^2} 
	+\frac{\beta}{8 \lambda}\right) \right]^{1/2} \right\}\,, 
\label{mhsigma}
\end{equation} 
where the plus sign refers to the heavier eigenstate. Assuming $f^2>0$ and  the
$SO(5)$ explicit breaking to be small (${|\beta|}/{4\lambda}\ll1$) , which can be implemented applying  't Hooft's naturalness principle, the masses of the Higgs and the $\sigma$ particle are given by:
\begin{equation}
m_{h}^2=~ 2\beta v^2+O\left(\frac{\beta}{4\lambda}\right)\,,\\
\end{equation}
\begin{equation}
m_{\sigma}^2=~ 8\lambda f^2 + 2\beta (3 f^2 - v^2) +O\left(\frac{\beta}{4\lambda}\right)\,.
\end{equation}
  These results show that, at variance with the SM case, in the regime of small soft $SO(5)$ breaking the mass of the Higgs and its
 quartic self-coupling are controlled by two different parameters $\beta$ and $\lambda$ respectively. 
 This is consistent with the PNGB nature of the Higgs boson whose mass should now appear protected from growing in the strong interacting regime of the theory,
 that is recovered in the limit $m_\sigma\gg f$, or equivalently $\lambda\gg 1$ (this connection has been demonstrated explicitly in Refs.~\cite{Feruglio:2016zvt,Gavela:2016vte}, showing that the ensuing effective operators are those usually considered in the literature in the context of composite Higgs models.).
 Only the $\sigma$ mass would increase in this limit.
 The hierarchy problem for the Higgs particle mass has then been replaced by a sensitivity of the $\sigma$ particle to heavier scales.
The expression for $m_h$ shows that the value of the $\beta$ parameter for small $\beta/4\lambda$ is expected to be 
$\beta\sim {m_h^2}/{2v^2}\sim 0.13$.

In a  second step of our analysis of the PQ symmetry, we will not use a linear sigma model but generalize instead the analysis to the large variety  of non-linear realizations of the Goldstone Higgs  existing in the literature~\cite{Carena:2014ria, Gavela:2016vte} (equivalent to the renormalizable linear sigma model with the $\sigma$ field integrated out), which widely differ in their exotic fermionic spectra.         
Only the third generation of SM quarks will be explicitly discussed, but the generalization to three light families is straightforward.

On the phenomenological arena, we will determine the implications of the minimal PQ implementation described  for axion-photon-photon couplings. The effective coupling $g_{a\gamma\gamma}$ is defined as   
\beq
 \delta\L_a \supset  - \frac{1}{4}g_{a\gamma\gamma} \,a\,F_{\m\n}\tilde{F}^{\m\n}\,\sim\, -g_{a\gamma\gamma}\,a(x)\,\mathbf{E}\cdot \mathbf{B}\,,
 \label{gagammagamma-def}
 \eeq
\beq\label{agammagamma_coupling}
g_{a\gamma\gamma}= \frac{m_a}{\text{eV}}\frac{2.0}{10^{10}\text{GeV}}\left( \frac{E}{N}-1.92(4)\right)\,,
\eeq
In these equations, $E$ and $N$ denote respectively the electromagnetic and color anomaly coefficients for a given fermion content $\Psi$,
\bea
N&=&\sum_\Psi \Big(\beta(\Psi_L)-\beta(\Psi_R)\Big)\,T(C_\Psi)\,, \nn\\
E&=&\sum_\Psi\Big(\beta(\Psi_L)-\beta(\Psi_R)\Big) \, \mathcal{E}_\Psi^2\,, \label{NE}
\eea
where $\beta(\Psi_L)$ and $\beta(\Psi_R)$ are respectively the PQ charges of the left-handed and right-handed components of a given fermion representation $\Psi$ (only the chiral differences are physical and relevant), $\mathcal{E}_\Psi$ is the $U(1)_{\text{em}}$ electromagnetic charge,  and  the color factor is given by $T(\mathcal{C}_\Psi)\delta^{ab}=\mathrm{Tr}\,T_\Psi^a T_\Psi^b$, where $\{a,b\}$ are color indices and $T_\Psi$ denotes the generator in the given color representation. 
 The ratio $E/N$ has been recently computed for different representations of exotic fermions in the standard ``invisible axion'' models in Refs.~\cite{DiLuzio:2016sbl, DiLuzio:2017pfr}.

In Sect.~2 we construct the PQ-invariant renormalizable model, exploring in all generality the possible extensions of its spectrum by scalar singlets and determining the ensuing constraints. In Sect.~3 we focus on the minimal case in which a single scalar singlet $S$ is added to the spectrum for the renormalizable model; in Sect.~4 we consider analogous PQ extensions for a plethora of non-linear realizations of a Goldstone Higgs with partial compositeness present in the literature. The different possibilities for charging the fermionic sector will be also discussed and the phenomenological predictions will be obtained. In Sect.~5 we conclude.


\section{A  renormalizable model:  the linear sigma model}
\label{Model}

 Following Ref.~\cite{Feruglio:2016zvt},  consider the minimal SO(5) linear $\s$-model,  with the symmetry softly broken to SO(4) and fermionic content given by:
  \bea
\mathcal{L}_{ferm}
   = &\bigg\{ &\bar{q}_{L} i\Ds  q_L + \bar{t}_R i\Ds  t_R 
   + \bar{\psi}^{(5)}i\Ds\psi^{(5)} + \bar{\psi}^{(1)}i\Ds\psi^{(1)}
   \nn \\
     &&-\Big[
    \bar{\psi}^{(5)}_L M_5 \psi^{(5)}_R + \bar{\psi}^{(1)}_L M_1 \psi^{(1)}_R+ 
     y_1\,\bar{\psi}_{L}^{(5)} \phi \,\psi_{R}^{(1)}+y_2\,\bar{\psi}_{R}^{(5)} \phi \,\psi_{L}^{(1)} 
   \nn \\
   &&+ \,\Lam_1\paren{\bar{q}_L{\spurion_{2\times5}}}\psi_R^{(5)}		          						      
    + \Lam_2 \,\bar{\psi}_L^{(5)} \paren{\spurion_{5\times1} t_R} + \Lam_3 \,\bar{\psi}_L^{(1)} t_R  +h.c.\Big]\,\,\bigg\}\nn \\
  + && \hspace{-0.7cm}\bigg\{ \psi \rightarrow \psi',\quad t_R\rightarrow b_R,\quad (M,\, \Lambda\,, y)_i \rightarrow  (M',\, \Lambda'\,, y')_i\,  \bigg \}\,,
\label{SO5Lag}
     \eea
where $q_L$, $t_R$ and $b_R$ denote respectively the SM doublet and singlet fermions of the third generation, and  $\phi$  is a  $SO(5)$ scalar five-plet  which contains the Higgs field.  $\psi^{(5)}$ and  $\psi^{(1)}$ denote exotic fermions in  the fundamental and singlet representations of $SO(5)$, respectively.\footnote{This model will be denoted MCHM$_{5-1-1}$  (minimal composite Higgs model $ 5-1-1$) in a notation that keeps track of the $SO(5)$ representation in which the exotic partners of the SM quark doublet, top and bottom are embedded, see  Sect.~\ref{section_models}. }   
 The ensemble of exotic fields can be decomposed in terms of $SU(2)_L$ eigenstates as:
\bea \label{SU2_decomposition}
&\psi^{(5)}\,\sim\,(X^{(5)},Q^{(5)},T^{(5)}) \quad ,\quad \psi^{'(5)}\,\sim\,(Q^{(5)\prime},X^{(5)\prime},B^{(5)})\,, \nn\\
&\psi^{(1)}\,\sim\,T^{(1)}\quad,\quad\psi^{'(1)}\sim B^{(1)}\,, \nn\\
&\phi = \left( H^T, \tilde{H}^T,\sigma\right)^T\,,
\eea
where $T$, $B$ and $\sigma$ are singlets under $SU(2)_L$, while all other fields are $SU(2)_L$ doublets. 
 Tab.~\ref{Table_SMgauge_decomposition} shows the SM charges for these fields, as well as for other fermion representations to be considered later on.  It shows as well 
the charges of various $SO(5)$ representations under the global group $U(1)_X$, which is customarily added to the global symmetry group to ensure correct hypercharge assignments for the SM fermions, with  the pattern of spontaneous symmetry breaking given by 
 \beq
 SO(5) \,\times\,U(1)_X \rightarrow SO(4) \,\times U(1)_X \approx SU(2)_L \,\times\, SU(2)_R \,\times\,U(1)_X\,.
 \eeq
The hypercharge Y corresponds then to a combination of the $U(1)_X$ and $SU(2)_R$ generators (denoted respectively by $X$ and $\Sigma_R^{(3)}$) given by 
\beq
Y=\Sigma_R^{(3)}+X\,.
\label{hypercharge}
\eeq
Two $U(1)_X$ charge values turn out to be compatible with SM hypercharge assignments: $2/3$ and $-1/3$.

The heavy -exotic- mass eigenstates result from diagonalizing the mass terms containing $\psi^{(5)}$ and  $\psi^{(1)}$,  shown in the 2nd and 3rd lines in Eq.~(\ref{SO5Lag}). The latter   describe the most general $SO(5)$-invariant mass terms, but for two of them which break  $SO(5)$ explicitly and softly: those proportional to  $\Lambda_1$, $\Lambda_2$  (plus their primed counterparts for the bottom sector). 
  \begin{table}[t] \footnotesize
\centering
\begin{tabular}{|  >{$}c<{$}|  >{$}c<{$} |   >{$}c<{$} |  >{$}c<{$} |  } 
\hline
&SO(5)\,\times\,U(1)_X &SU(3)_C\,\times\,SU(2)_L\,\times\,U(1)_Y  & q_{\text{EM}} \\ 
\hline\hline
\psi^{'(1)}&(1,\,-1/3) & B^{(1)} = (3,\, 1,\,-1/3) &-1/3  \\
\hline
\psi^{(1)}&(1,\,2/3)&T^{(1)} = (3,\,1,\,2/3) & 2/3\\
\hline
\multirow{3}{*}{$\psi^{'(5)}$} &\multirow{3}{*}{$(5,\,-1/3)$} &  Q^{(5)} = (3,\, 2,\,1/6) & -1/3,\,2/3\\ 
					&& X^{(5)} = (3,\, 2,\,-5/6) &-1/3,\,-4/3\\ 
					&&B^{(5)} = (3,\, 1,\,-1/3)&-1/3   \\
					\hline
\multirow{3}{*}{$\psi^{(5)}$}&\multirow{3}{*}{$(5,\,2/3)$} &  Q^{(5)} = (3,\, 2,\,1/6)&-1/3,\,2/3 \\ 
					&& X^{(5)} = (3,\, 2,\,7/6) &2/3,\,5/3\\ 
					&&T^{(5)} = (3,\, 1,\,2/3) &2/3 \\
					\hline
\multirow{6}{*}{$\psi^{(10)}$} &\multirow{6}{*}{$(10,\,2/3)$} & Q^{(10)} = (3,\, 2,\,1/6)&-1/3,\,2/3 \\ 
					&& X^{(10)} = (3,\, 2,\,7/6)&2/3,\,5/3 \\  
					&& V^{(10)} = (3,\, 3,\,2/3) &-1/3,\,2/3,\,5/3\\ 
					&& W^{(10)} = (3,\, 1,\,5/3)&5/3\\ 
					&& B^{(10)} = (3,\, 1,\,-1/3)& -1/3 \\ 
					&&T^{(10)} = (3,\, 1,\,2/3) &2/3 \\
\hline
\multirow{6}{*}{$\psi^{(14)}$} &\multirow{6}{*}{$(14,\,2/3)$} & Q^{(14)} = (3,\, 2,\,1/6)&-1/3,\,2/3 \\ 
					&& X^{(14)} = (3,\, 2,\,7/6)&2/3,\,5/3 \\  
					&& V_1^{(14)} = (3,\, 3,\,5/3) &2/3,\,5/3,\,8/3\\ 
					&& V_1^{(14)} = (3,\, 3,\,2/3)&-1/3,\,2/3,\,5/3\\ 
					&& V_3^{(14)} = (3,\, 3,\,-1/3)& -4/3,\,-1/3,\,2/3 \\ 
					&&T^{(14)} = (3,\, 1,\,2/3) &2/3 \\
\hline
\end{tabular}
\caption{$SO(5)$ fermion representation content in terms of SM quantum numbers. 
The last column shows the $U(1)_{\text{EM}}$ electromagnetic charges, used to compute the electromagnetic anomaly $E$.} \label{Table_SMgauge_decomposition}
\end{table}
$\spurion_i$ are dimensionless matricial coefficients.  The quantities $(\Lambda_1\spurion_{2\times5})$ and $(\Lambda_2\spurion_{5\times1})$ act like spurions breaking the global $SO(5)$ symmetry;  they are  expected to be small compared to the overall $SO(5)$ scale $\Lambda_s$ and thus to  the overall scale of the exotic fermion masses $M_i$. They induce electroweak symmetry breaking through their one-loop contribution to the Coleman-Weinberg potential: they generate 
the electroweak scale $v$ and provide a mass for the Higgs particle which is small compared to $\Lambda_s$; they are thus expected to be in general about one or two orders of magnitude smaller than $\Lambda_s$, e.g. in the hundreds of GeV-TeV range.  The $\Lambda_i$ terms are also essential in generating light fermion masses. 
Indeed, in  the partial compositeness paradigm the direct SM Yukawa coupling is forbidden by the global symmetry, and a chain of interactions mediated by heavier fields is required in a seesaw-like structure.  For instance, in the renormalizable model at hand   the dominant contribution to the $\bar{t}_Lt_R$ mass term  corresponds schematically to\footnote{There are subleading contributions to the light fermion masses  which do not depend on $\Lambda_3$, see Ref.~\cite{Feruglio:2016zvt}. They could lead to PQ solutions other than those considered below, which focus on the leading option shown in Eq.~(\ref{tbmass}).}
\begin{equation}
q_L \underset {\Lambda_1}{\longrightarrow}  Q_R \underset {M_5}{\longrightarrow}  Q_L \underset {y_1\langle\tilde{H}\rangle}{\longrightarrow} T_R^{(1)} \underset {M_1}{\longrightarrow} T_L^{(1)}\underset {\Lambda_3}{\longrightarrow} t_R\,, 
\label{chain_renormalizable}
\end{equation}
and analogously for  the bottom mass, leading to
 \begin{equation}
 m_t \sim y_1\frac{\Lambda_1 \Lambda_3}{M_1 M_5}\, v\,,\qquad\qquad
 m_b \sim y_1'\frac{\Lambda_1' \Lambda_3'}{M_1' M_5'}\, v\,. 
 \label{tbmass}
\end{equation}    
It is easy to verify that the renormalizable  Lagrangian in Eq.~(\ref{SO5Lag}) is {\it not}  PQ invariant. Note nevertheless that --as customary in most partial compositeness realizations-- exotic vectorial fermions are by construction present, which suggests the possibility of a PQ-invariant extension ``\`a la KSVZ''. The novel ingredient inbuilt in partial compositeness scenario is precisely the chain of interactions needed to generate fermion masses via Yukawa couplings with exotic heavy fermions. This constraint will strongly reduce the  freedom in the relative choice of PQ charges for the exotic fermions, and increase predictivity.    We explore next the possible minimal PQ invariant extensions of the linear $SO(5)/SO(4)$ sigma model~\cite{Feruglio:2016zvt} with its original fermion content as shown  in Eq.~(\ref{SO5Lag}), enlarging only its scalar sector.

 The need of a PQ scale much higher than the overall $SO(5)$ scale $\Lambda_s$ suggests to introduce it through the vev of a scalar field (or fields) $S$ ($S_i$), singlet under the SM and under the global $SO(5)$ symmetry and charged under PQ, e.g.,
\begin{equation}
 S(x)=\frac{f_a+\rho(x)+ia(x)}{\sqrt2}\,,
 \label{Sdefinition}
\end{equation} 
where $f_a$ sets the PQ scale as the vacuum expectation value (vev) of $S$, $\langle S \rangle \sim f_a$, 
the real field $\rho(x)$ is expected to be heavy\footnote{Of $\mathcal{O}(f_a)$. Its dynamics is thus not relevant at low energies; in particular, $\rho(x)$ can be considered integrated out of the spectrum for some cases discussed below with very high $f_a$ scale.} and the  imaginary part is to be identified with the axion $a(x)$, which is massless at the classical level.

With only one five-plet scalar in the spectrum of the model, it is not possible to give PQ charge to this field, as an even number of components is needed for it to be charged.\footnote{A ten-component $SO(5)$ multiplet must be built (e.g. out of two scalar five-plets), in order for it to be PQ charged and still contain all four components of the Higgs doublet: $H$ and its conjugate $\tilde{H}$. We defer this alternative extension path to a future exploration.} The fermions instead can easily acquire PQ charges.  There are many options for selecting which fermionic mass parameters in the Lagrangian are promoted to dynamical fields so as to implement the PQ symmetry. We will first derive general constraints which are valid  for the case in which either just one or more than one  scalar singlet would be added. 

Extending the model spectrum by only singlet scalars $S_{i}$, the most general renormalizable model would correspond to promoting to independent dynamical fields {\it all} $M_i$ and $\Lambda_i$ (that is, $M_1$, $M_5$ and $\Lambda_{1,2,3}$) parameters in the Lagrangian in Eq.~(\ref{SO5Lag}),
 \bea\label{promoting}
 &M_i&\to \kappa_i S_{Mi}\,,\\
 &\Lambda_i&\to \lambda_i S_{\Ll i}\,,\label{promoting2}
 \eea 
where $\kappa_i$ and $\lambda_i$ are constants and $S_i$ are independent fields with generically $\langle S_i\rangle \sim f_a$.  $\kappa_i$ and $\lambda_i$ are necessarily small 
if the physical value of the exotic fermion masses is $\sim \Lambda_s$. This is 
technically natural  in the 't Hooft sense if the small  $\kappa_i$ and $\lambda_i$ values are protected by a symmetry: we  find that indeed   all  models explored with $\mathcal{O}(\Lambda_s)$ fermion masses are protected by chiral symmetries under which the fermions transform but not the scalars. Alternative setups with $\kappa_i$ and $\lambda_i$ of $\mathcal{O}(1)$ are possible a priori by allowing some of the $M_i$ and $\Lambda_i$ values to be of $\mathcal{O}(f_a)$, so as to cancel the $f_a$ dependence between the numerator and denominator in Eq.~(\ref{tbmass});  this may be a safe option from the point of view of naturalness and stability of the scalar potential only if  the $\Lambda_i$ parameters  involved do not correspond to terms breaking the global symmetry and if the choice is protected at the quantum level, which in practice points to promoting singlet fields, see further below. Nevertheless, although the contributions of heavy $SO(5)$ singlet fermions will not destabilize the Higgs mass,  it could affect and destabilize the value of the overall scale $f$ itself: this issue cannot be settled without a specific one-loop analysis of the potential which is beyond the scope of this paper. In the absence of such analysis, we choose not to discard this type of solutions below. 
For the results in the rest of this section, all $\kappa_i$ and $\lambda_i$ will be general arbitrary parameters. 

As previously done for fermions, see Eq.~(\ref{NE}), we refer to the PQ charges of scalars as $\beta (\phi)$, where $\phi$ is a generic scalar.
For instance,  $\beta(S_{Mi })$ and $\beta(S_{\Lambda i})$ will denote the PQ charges of the fields resulting from promoting to dynamical variables the fermionic mass parameters, as indicated in Eqs.~(\ref{promoting}) and \eqref{promoting2}.
The following general set of constraints follows for the top sector\footnote{We allow here  top and bottom quarks charged under PQ.  Alternatively, it would be possible to charge any of the other two light quark generations. If mixings among light families are taken into account, the generalization to three families may imply charging under PQ all three fermion generations, depending on the model. } in order to achieve PQ invariance:    
\begin{align} 
\beta(\psi^{(5)}_R)&=	\beta(q_L) - \beta(S_{\Lambda1})\,, \label{psiupL} \\ 
\beta(\psi^{(5)}_L)& =	\beta(q_L)  - \beta(S_{\Lambda1})                       + \beta(S_{M5})\,,                                     
\\
 \beta(\psi^{(1)}_R)& =	\beta(q_L) - \beta(S_{\Lambda1})+\beta(S_{M5 }) \,, \label{chiupR}    \\
  \beta(\psi^{(1)}_L)&=	\beta(q_L) - \beta(S_{\Lambda1})+\beta(S_{M5 }) +\beta(S_{M1 })\,,  \label{chiupL} \\
                \beta(t_R)&=	\beta(q_L) - \beta(S_{\Lambda1})   -\beta(S_{\Lambda3}) +\beta(S_{M5 })+ \beta(S_{M1 })  \,,             \label{tR}                   \\
      \quad & \quad \nn\\
\beta(S_{\Lambda2})&=	\beta(S_{\Lambda3})-\beta(S_{M1}) \label{Lambda2}\,,\\
0&=	\beta(S_{M1})+\beta(S_{M5})\,.\label{last}
\end{align}
The last two constraints, Eqs.~(\ref{Lambda2}) and (\ref{last}), are respectively those stemming from the Lagrangian terms proportional to $\Lambda_2$ and $y_2$.  They are special in the sense that the presence in the Lagrangian of the parameters  $\Lambda_2$, $\Lambda_2'$, $y_2$, $y_2'$ is not strictly necessary, as they only induce subleading contributions to the top and bottom masses~\cite{Feruglio:2016zvt}. 
The absence of some or all of them 
may be protected from radiative instability by symmetries; furthermore, the PQ symmetry itself can guarantee their absence at all orders depending on its implementation, as illustrated further below. Would they be absent in a given model,  the number of constraints implied by PQ charge conservation would correspondingly decrease and the parameter space would  be enlarged in consequence. 

 The set of Eqs.~(\ref{psiupL})-(\ref{last}) is accompanied by the analogous constraints stemming from the bottom sector, obtained from the above via the replacement
\begin{equation}
  \bigg\{ \psi \rightarrow \psi',\quad t_R\rightarrow b_R,\quad (M,\, \Lambda \,)_i \rightarrow  (M',\, \Lambda'\, )_i\,  \bigg \}\,.
  \label{top-bottom}
  \end{equation}
They amount in total to 14 equations with 21 free parameters, which leaves much freedom in the choice of dynamical parameters and PQ charges.  The individual PQ charges of left-handed and right-handed fields  are not physical {\it per se}: the only quantities relevant  for the computation of the color and electric anomalies are  the chiral differences, 
\begin{equation}\label{chiralcharge}
\LRdel \Psi \equiv \beta(\Psi_L)-\beta(\Psi_R)\,,
\end{equation}
where $\Psi$ denotes generically a fermion. Here and in what follows {\it we will refer to fermions with a non-vanishing $\LRdel{\Psi}$ as having a chiral PQ charge.}
While this definition for the vectorial fermions  $\psi^{(1)}$ and $\psi^{(5)}$ and their primed counterparts is straightforward,  the chiral PQ charges of the SM top and bottom fields (whose left and right components are not directly coupled in the Lagrangian Eq.~(\ref{SO5Lag})) will be defined as 
\begin{equation}\label{chiraltbcharge}
\LRdel t \equiv \beta(q_L)-\beta(t_R)\,, \qquad \qquad \LRdel b \equiv \beta(q_L)-\beta(b_R)\,.
\end{equation}
Note that charging $q_L$ under PQ  implies charging both the top and bottom left-handed quarks, but this does not necessarily imply $\LRdel t\ne 0$ and/or $\LRdel b\ne 0$. 
{\it As for fermions only the chiral PQ differences are physical, 
$\LRdel t$ and $\LRdel b$ will be retained as the physically relevant quantities to analyze the top and bottom sectors.  
}

The fermionic PQ chiral charges can be expressed in terms of the PQ charges of the scalar fields:
\begin{equation}
\begin{aligned} \label{Chiral_Differences}
&\LRdel \psi^{(5)}=\beta(S_{M5})\,,\qquad \LRdel \psi^{(1)} =\beta(S_{M1})\,,  \\
& \LRdel t=\beta(S_{\Lambda1})+\beta(S_{\Lambda3})-\beta(S_{M1})-\beta(S_{M5})\,,
\end{aligned}
\end{equation}
 plus those for the bottom sector obtained via the replacement in Eq.~(\ref{top-bottom}).
 The quantities $E$ and $N$ can then be expressed  as general functions of the fermionic PQ chiral charges, resulting in 
\bea
\label{general_EN}
E&=& \,\frac 13 \bigg[ 38 \LRdel \psi^{(5)}+23 \LRdel \psi^{'(5)}+4 \LRdel \psi^{(1)}+\LRdel \psi^{'(1)}+4 \LRdel t + \LRdel b \bigg] \,,\\
N &=&\frac12 \bigg[5\, \LRdel \psi^{(5)} + 
     5\,\LRdel \psi^{'(5)}+ \LRdel \psi^{(1)}+\LRdel \psi^{'(1)}+ \LRdel t + \LRdel b\bigg]\,.
\eea
 Using Eq.~(\ref{Chiral_Differences}), these equations allow to express the ratio $E/N$  in terms of the PQ charges of the scalar fields,
\begin{equation} \label{general_E/N}
\frac{E}{N} = \frac{2}{3}\frac{34 \beta(S_{M5}) + 22 \beta(S_{M5'})+ 4\beta(S_{\Lambda 1})+4\beta(S_{\Lambda3})+\beta(S_{\Lambda 1 '})+ \beta(S_{\Lambda3'})}{4 \beta(S_{M5}) + 4 \beta(S_{M5'})+ \beta(S_{\Lambda 1})+\beta(S_{\Lambda3})+\beta(S_{\Lambda 1 '})+ \beta(S_{\Lambda3'})}\,.
\end{equation} 
The number of possible PQ invariant setups reduces  if some of the dimensionful parameters are not promoted to dynamical fields, or if several singlet scalar fields are identified among themselves.  In fact, when more than one extra scalar singlet is present, relations among their charges may need to be established (for instance through couplings in the scalar potential) if we would only wish to implement one PQ symmetry, and thus a single axion, instead of a plethora of axial $U(1)$ symmetries with their corresponding GBs.   As stated, from now on we focus on  analyzing the minimal addition of a single scalar singlet $S$. Note that in this case 
each charge $\beta(S_i)$ is either 0 or $\pm \beta(S)$, depending on whether a coupling is promoted to $S$ or $S^\dagger$.

\section{Extension by only one scalar singlet $S$: the renormalizable model}\label{section_our_model}
      
In order to gain perspective,  two extreme setups with only one extra scalar singlet $S$ will be explored in detail for the renormalizable model presented in the previous section:  i) the case in which only one fermion is chirally charged under PQ, and  ii)  the option in which all fields and couplings are  allowed to be freely and arbitrarily charged. For both cases,  the values of $E/N$ corresponding to the maximum and minimum $|g_{a\g\g}|$ attainable will be evaluated. This will allow a comparison with the predictions of recent updated analysis of the standard KSVZ and DFSZ theories. We will first develop  the discussion assuming implicitly that all the exotic fermion sector has masses of $\mathcal{O} (\Lambda_s)$, to discuss next which ones among the solutions could {\it a priori} allow instead $\mathcal{O} (f_a$) fermion mass parameters.

\subsection{Only one exotic fermion representation chirally charged under PQ} 
  
The rationale for considering  first only one exotic fermion charged is simplicity to illustrate the procedure, and also to allow an easy comparison with recent updated analyses of the standard KSVZ and DFSZ theories~\cite{DiLuzio:2016sbl, DiLuzio:2017pfr}, which start by extending the SM fermionic sector by only  one exotic fermion. There is otherwise no special advantage or economy in preventing more than one fermion (among those {\it required} by the composite Higgs model) to be chirally charged under PQ, and this general option will be explored later on.

All  PQ-invariant solutions of the composite Higgs model that assume only one heavy exotic fermion charged under PQ require either $y_2 = 0$ or $y_2' = 0$ in order to implement PQ invariance, as otherwise the PQ charges of both exotic fermions are linked, see  Eq.~(\ref{last}). 
Tab.~\ref{TableOnlyOneS} displays different examples of this kind, together with their corresponding phenomenological predictions for axion-photon interactions.  
     
A simple example is to charge under PQ 
only $S$ and  the $SO(5)$ fermion singlet $\psi^{(1)}_L$ , by promoting  the mass $M_1$   and $ \Lambda_3$ to dynamical variables:
 \bea
 &M_1&\to \kappa_1 S\,,\nn\\
  &\Lambda_3&\to \lambda_3 S\,.
 \label{natural-sol}
 \eea
$M_1 = \kappa_1 f_a$ is thus generated dynamically, together with the coupling responsible for the decay of the exotic quarks into SM quarks $\Lambda_3 = \lambda_3 f_a$. The application of this prescription to Eq.~(\ref{SO5Lag}) renders a PQ symmetric Lagrangian, with $\psi^{(1)}_L$ charged under PQ and $\psi^{(1)}_R$ uncharged.\footnote{An analogously economical and natural model  consists in charging instead ${\psi^{(1)\prime}}$, with $M_1'$ and $\Lambda_3'$ becoming dynamical.}
Furthermore, the condition $y_2=0$ is protected from quantum instabilities by the PQ symmetry itself.   In turn, the very small values required for the parameters $\kappa_1$ and $\lambda_3$ (assuming $M_i$ masses not higher than 10 or 100 TeV as usual in composite Higgs models) are natural in 't Hooft sense~\cite{tHooft:1979rat}. Indeed, they are protected by a chiral symmetry: that in which  only $\psi^{(1)}_L$ transforms,  and neither $S$ nor any other  field does.  This is a pattern which will hold for basically all cases to follow: as the Lagrangian parameters which are being promoted to dynamical fields are fermionic mass parameters, their absence should be expected to be related to new chiral symmetries, rendering technically natural the choice of small values for the $\kappa_i$ and $\lambda_i$ parameters.

An alternative simple solution also with only one exotic heavy field  charged under PQ is given by  promoting to dynamical fields the parameters relevant for the 
five-plet fermionic field $\psi^{(5)}$, 
 \bea
  &M_5&\to \kappa_5 S\,,\nn\\
  &\Lambda_1&\to \lambda_1 S\,.
  \eea
This solution can be realized charging under PQ only $\psi^{(5)}_R$ and $S$, which requires $y_2=0$.   Again, the small values phenomenologically required in this case for $\kappa_5$ and $\lambda_1$ are technically natural, as in their absence the Lagrangian acquires a chiral symmetry under which only $\psi^{(5)}_R$ would transform. The main contrast with the previous example is that here a parameter associated to a symmetry-breaking term, $\Lambda_1$, has been promoted to dynamical field. Intuitively, it is expected that its value should be smaller than those corresponding to $SO(5)$ invariant terms, such as the $M_i$ diagonal mass terms or the $\Lambda_3$ coupling. In other words, a dynamically promoted $\Lambda_1$ requires a slightly stronger adjustment for the $\lambda_1$ parameter than in the previous example (e.g. by one or two orders of magnitude). For this reason it may be preferred to avoid solutions which promote $\Lambda_1$ or $\Lambda_2$ to dynamical fields,  although strictly speaking they are still technically natural solutions and thus valid ones.  All options in which $M_5$ is promoted to a dynamical field require $\Lambda_1$ or $\Lambda_2$ to be also dynamical, except if a SM quark is simultaneously allowed to acquire a chiral PQ charge,\footnote{Analogous putative solutions with only a singlet exotic fermion ($M_1$) plus a SM fermion chirally charged under PQ, and no dynamical $\Lambda_i$, are not possible because the contribution to the color anomaly cancels in that case, leaving the strong CP problem unsolved.} see Tab.~\ref{TableOnlyOneS}; this case belongs then to the class of solutions with more than one PQ-charged fermion to be discussed later on.
\begin{table}[h!]
\begin{center}
\begin{tabular}{  >{$}c<{$} | >{$}c<{$} | >{$}c<{$}  | >{$}c<{$} }
\hline
  $Exotic fermion charged$ &  $Couplings promoted$&$SM fermions charged $  & E/N_{\text{top}}\, (E/N_{\text{bottom}})\\[2mm]
  \hline
  \hline
$\multirow{4}{*}{$\begin{matrix} \LRdel\psi^{(1)}=\beta(S) \\ \\ M_1 \rightarrow \kappa_1 S \end{matrix}$} $
						&{\Ll_3 {\rightarrow}  \lambda_3 S}&-&$ \multirow{2}{*}{$\frac 83\quad \left(\frac{2}{3}\right)$}$\\\cline{2-3}
						& \begin{matrix} \Ll_1 \rightarrow \lambda_1 S \\ \Ll_2 \rightarrow  \lambda_2 S^\dagger \end{matrix}  &-& \\\cline{2-3}
						&\begin{matrix} \Ll_{1,\,3}\rightarrow \lambda_{1,\,3} \,S  \end{matrix} &\LRdel t=\beta(S)  & \\\cline{2-3}
						&\begin{matrix} \Ll_{1,\,2}\rightarrow \lambda_{1,\,2} \,S^\dagger \end{matrix} &\LRdel t=-2\beta(S)  & \\\cline{2-3}
\hline
\hline
$\multirow{9}{*}{$\begin{matrix} \LRdel\psi^{(5)}=\beta(S) \\ \\ M_5 \rightarrow \kappa_5 S \end{matrix}$} $
						&\Ll_1 \rightarrow  \lambda_1 \,S&-&$ \multirow{2}{*}{$\frac{76}{15} \quad \left(\frac{46}{15}\right)$}$\\\cline{2-3}
						& \begin{matrix} \Ll_{2,\,3} \rightarrow \lambda_{2,\,3}\, S  \end{matrix}  &-& \\\cline{2-4}
						&\begin{matrix} \Ll_{1,\,2,\,3}\rightarrow \lambda_{1,\,2,\,3}\, S\end{matrix} &\LRdel t=\beta(S)  &\frac{14}{3} \quad \left(\frac{8}{3}\right)\\\cline{2-4}
						&-&{\LRdel t=-\beta(S)}  &$ \multirow{2}{*}{$\frac{17}{3} \quad \left(\frac{11}{3}\right)$}$\\\cline{2-3}
						&\begin{matrix} \Ll_1\rightarrow \lambda_1 S^\dagger\,(\lambda_1 S) \\ \Ll_{2,\,3} \rightarrow \lambda_{2,\,3}\, S\, (\lambda_{2,\,3}\, S^\dagger) \end{matrix}&{\LRdel t=-\beta(S)}  & \\\cline{2-4}
						&\Ll_1 \rightarrow \lambda_1 S^\dagger  &\LRdel t=-2\beta(S)  &$\multirow{2}{*}{$\frac{20}{3} \quad \left(\frac{14}{3}\right)$}$ \\\cline{2-3}
						&\begin{matrix}  \Ll_{2,\,3} \rightarrow \lambda_{2,\,3} S^\dagger  \end{matrix} &\LRdel t=-2\beta(S)  & \\\cline{2-4}

						&\begin{matrix}  \Ll_{1,\,2,\,3} \rightarrow \lambda_{1,\,2,\,3}\, S^\dagger  \end{matrix} &\LRdel t=-3\beta(S)  &\frac{26}{3} \quad \left(\frac{20}{3}\right) \\\cline{2-4}
\hline
\hline
$\multirow{3}{*}{$\begin{matrix} \text{None} \\ \\  \end{matrix}$} $
						&{\Ll_1 {\rightarrow}  \lambda_1 S}&\LRdel t=\beta(S)&$ \multirow{2}{*}{$\frac 83\quad \left(\frac{2}{3}\right)$}$\\\cline{2-3}
						& \begin{matrix} \Ll_{2,\,3} \rightarrow \lambda_{2,\,3} S\end{matrix}  &\LRdel t=\beta(S)& \\\cline{2-3}
						&\begin{matrix} \Ll_{1,\,2,\,3}\rightarrow \lambda_{1,\,2,\,3}\, S  \end{matrix} &\LRdel t=2\beta(S)  & \\\cline{2-3}
						\hline
\hline
\end{tabular}
\end{center}
\vspace{-0.5cm}\caption{Possible setups extending the spectrum of the renormalizable composite Higgs model by one singlet scalar S and allowing one or none of the exotic fermions to acquire chiral PQ charges. 
 Either the top or the bottom sector are considered at a time; the top sector is explicitly illustrated, while for the bottom sector ($M_i\to M_i',\,\Lambda_i\to\Lambda'_i$ with $\Lambda_i=0$) the $E/N$ values are shown within brackets.
 }\label{TableOnlyOneS}
\end{table}

\subsubsection{One exotic fermion charged, with mass parameters of $\mathcal{O} (f_a)$} 
While the bulk of the solutions is that discussed above with all fermion masses around the TeV range, we consider here a few particular additional solutions: those with still only one fermion chirally charged, although assuming now that it is a very heavy exotic one (and all SM fields uncharged).   This option is appealing in the sense that,  if such a large scale exists  in Nature, dimensionful dynamical parameters of the complete Lagrangian may tend to be of that order, if they correspond to terms invariant under the lower energy symmetries (the SM gauge group and the global $SO(5)$ symmetry in the case under discussion). In our Lagrangian Eq.~(\ref{SO5Lag}) $M_1$, $M_5$ and $\Lambda_3$  are of this kind, while only the terms proportional to $\Lambda_1$ and $\Lambda_2$ break $SO(5)$ (and analogously for  the primed counterparts of the bottom sector). 
Intuitively,  the dimensionful couplings that involve only singlet fields,  and are therefore insensitive to the $SO(5)$ structure, are expected to be of the order of the largest scale in Nature to which they can connect.  If that sector is decoupled from the non-singlet one, small parameters will not be required elsewhere either.   This point of view would select $M_1$ and $\Lambda_3$ as putative scales of $\mathcal{O} (f_a)$.   Indeed, among the solutions of the Lagrangian Eq.~(\ref{SO5Lag}) gathered in Tab.~\ref{TableOnlyOneS}: 
   
\begin{itemize}
    \item The solution in Eq.~(\ref{natural-sol}) with $M_1$ and $\Lambda_3$ dynamical  (and/or its primed counterpart) does not require any tuning of the parameters $\kappa_3$ and $\lambda_1$.  That is, both of them can be in this case of $\mathcal{O} (1)$, see nevertheless the caveats regarding the stability of the scale $f$ after Eq.~(\ref{promoting2}). The singlet and non-singlet sectors are disconnected, as the PQ symmetry itself forbids the presence of $y_2$ (or $y_2'$).
    The $f_a$ dependence cleanly cancels in the SM mass expressions in Eq.~(\ref{tbmass});  the Yukawa coupling $y_1$ (or $y_1'$) does not requires fine-tuning either. These models are depicted by blue lines in Fig.\ref{ma_ga_ParamSpace:sub1}: $E/N=8/3$ for the solution in the top sector (lower line), and $2/3$ for that in the bottom sector (upper line).
    \item In contrast, the algebraic solution in which  $M_5$ is the only  mass parameter which becomes dynamical (and large, with $\kappa_5\sim\mathcal{O}(1)$), with  $\psi_R^{(5)}$ and the SM fermion $q_L$ charged under PQ, would be possible  
    only at the unacceptable price of a very large Yukawa coupling $y_1\sim f_a M_1/(\Lambda_1 \Lambda_3)\gg 4\pi$,  well outside its perturbative range. 
    This is because in the expression for the light fermion masses,  Eq.~(\ref{tbmass}), no other mass parameter is large enough so as to compensate the $M_5\sim f_a$ dependence of the denominator.
   
    \item The solutions in which either $\Lambda_1$ or $\Lambda_2$ would be of $\mathcal{O} (f_a)$   (that is,  $\lambda_1 $ or $\lambda_2$  of $\mathcal{O} (1)$)   seem also unacceptable, for the naturalness reasons explained. From the sole point of view of the SM fermion masses in Eq.~(\ref{tbmass}) they could be acceptable, in particular those in which the $f_a$ dependence cancels between numerator and denominator. The question that would need clarification, though, is whether a large $\Lambda_1$ and/or $\Lambda_2$
would induce  inordinately large $SO(5)$-breaking terms in the  effective potential, rendering it unstable and spoiling the GB character of the Higgs field. Note that the electroweak scale $v$ and the Higgs mass must be ultimately proportional to the only $SO(5)$-breaking parameters of the model, $\Lambda_{1,2}$, unless ad-hoc fine-tunings are implemented in the scalar potential. In the absence of a satisfactory justification, it is safer to disregard these solutions  with $\Lambda_{1,2}\sim\mathcal{O}(f_a)$ (in contrast to the case in which they are much smaller, as discussed in the previous subsection).

    \end{itemize}
A general question raised by very heavy fermions  is their compatibility with the phenomenological constraints on the $S$, $T$ and $U$ parameters and other electroweak precision tests. Perfect vector-like fermions (with identical masses) do not contribute to $S$, $T$ and $U$ and a large overall scale is not an issue then. Their contributions when non-degenerate are suppressed by the vector-like masses $M_i$, but enhanced by the $\Lambda_i$ parameters. What really matters then is the mixing, which is again set by $\Lambda_i/M_i$ ratios. It follows that  the  preferred solution identified in the case of fermion masses of $\mathcal{O} (f_a)$,  which involves only the singlet $SO(5)$ fermion, Eq.~(\ref{natural-sol}), 
could be both natural and not subject to extra phenomenological tensions, up to the question of whether the scale $f$ may be destabilized in this scenario. 
     
\vspace{0.2cm}
All the above considerations about $\mathcal{O} (f_a)$ exotic fermion mass parameters will apply as well to the various different composite Higgs models discussed further below. In any case, the numerical predictions for the $E/N$ factor which determine the strength of axion-photon-photon couplings are independent of the  values of the exotic heavy fermion mass parameters; the sole criteria to discriminate among models with different fermionic scales is the conceptual one discussed here.

\subsection{Only one SM fermion chirally charged under PQ} 
  
In the case of traditional KSVZ invisible axion models, the options with just one fermion charged under PQ necessarily imply that the fermion is an exotic one,  because in these models  the SM fermions cannot acquire PQ charges, a fact that follows from the SM Yukawa couplings, which induce the same constraint on fermion couplings as that required by gauge hypercharge anomaly cancellation. For the partial fermion compositeness paradigm instead, as  there are no  Yukawa couplings linking the left and right components of SM fermions but only Yukawa couplings involving the exotic heavy fermions, SM fermions can be chirally charged under PQ. This can be easily understood from the chain of couplings required to generate fermion masses, Eq.~(\ref{chain_renormalizable}): by promoting to dynamical fields some of the exotic mass parameters $\Lambda_i$, the PQ charge of the left and right components of a given SM fermion do not need to coincide. 
      
PQ-invariant solutions of the composite Higgs model in which the only PQ-charged fermion is a SM one are also shown in Tab.~\ref{TableOnlyOneS}. They correspond to either $\LRdel t\ne0$ or $\LRdel b\ne0$.  These solutions do not require $y_2$ or $y_2'$ to vanish to enforce PQ invariance. 
Note that because of the chiral character of SM fermions, the illustration would be slightly different if the analysis was developed in terms of ``only one fermion representation'', as in that case charging for instance  $q_L$ would give additional results, but this would correspond to considering {\it two} chiral differences, $\LRdel t\ne0$ {\it and} $\LRdel b\ne0$.
\footnote{This requires to charge as well $\Lambda_1$ and $\Lambda_1'$, resulting  $\LRdel t= \LRdel b=\pm \beta(S)$ and $E/N=5/3$.  All cases are anyway included further below when allowing all fermions to get simultaneously arbitrary PQ charges.}

\begin{figure}[t]
\centering
\begin{subfigure}{.85\textwidth}
  \centering
  \includegraphics[width=.95\linewidth]{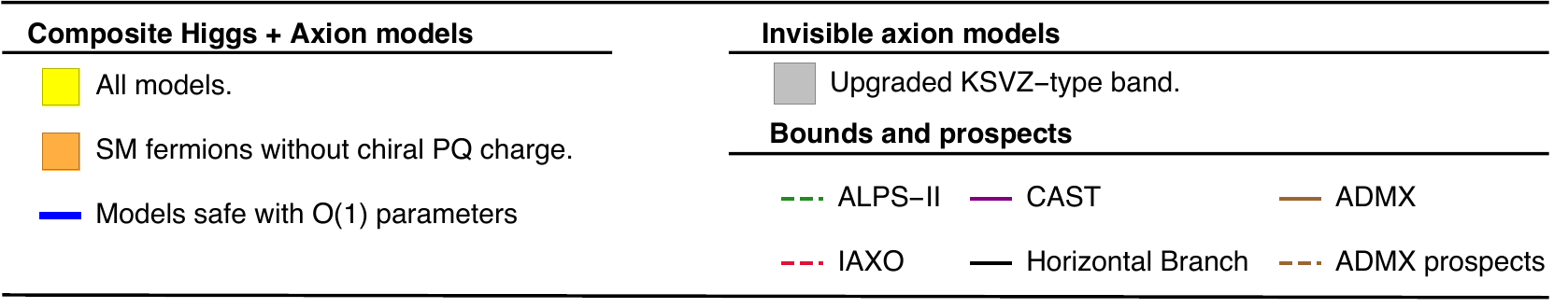}
  \end{subfigure}%
\vspace{.3cm}
\subcaptionbox{ In yellow, $|g_{a\gamma\gamma}|$ values allowed when only one fermion representation is chirally charged amongst the ensemble of those for exotic and SM fermions, in the renormalizable composite Higgs model. 
 The subset of models in which only the exotic fermions are PQ-chirally charged (orange hatched) spans the same maximal region in this case.
The grey area corresponds to the updated~\cite{DiLuzio:2016sbl} standard KSVZ prediction.
 \label{ma_ga_ParamSpace:sub1}
}%
  [.47\linewidth]{\includegraphics[width=.4\textwidth]{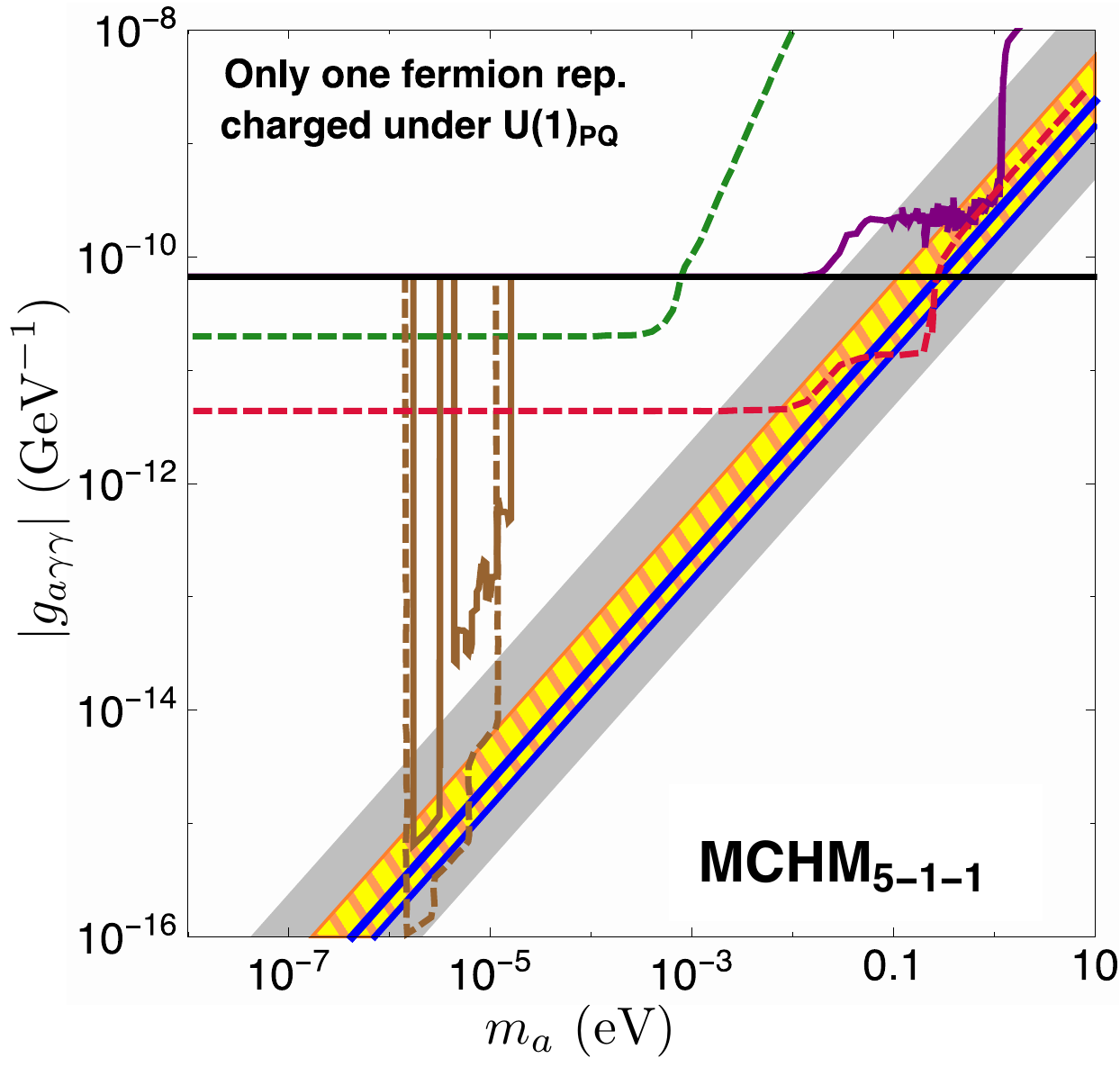}} 
  \hspace{.5cm}
\subcaptionbox{ The yellow band indicates the allowed values of $|g_{a\gamma\gamma}|$ when more than one  fermion is allowed to be charged under $U(1)_{\mathrm{PQ}}$. The subset of models in which the SM fermions are not chirally charged under PQ is indicated by an orange band.  The grey band corresponds to the upgraded~\cite{DiLuzio:2016sbl} KSVZ prediction. \label{ma_ga_ParamSpace:sub2}}
  [.47\linewidth]{\includegraphics[width=.4\textwidth]{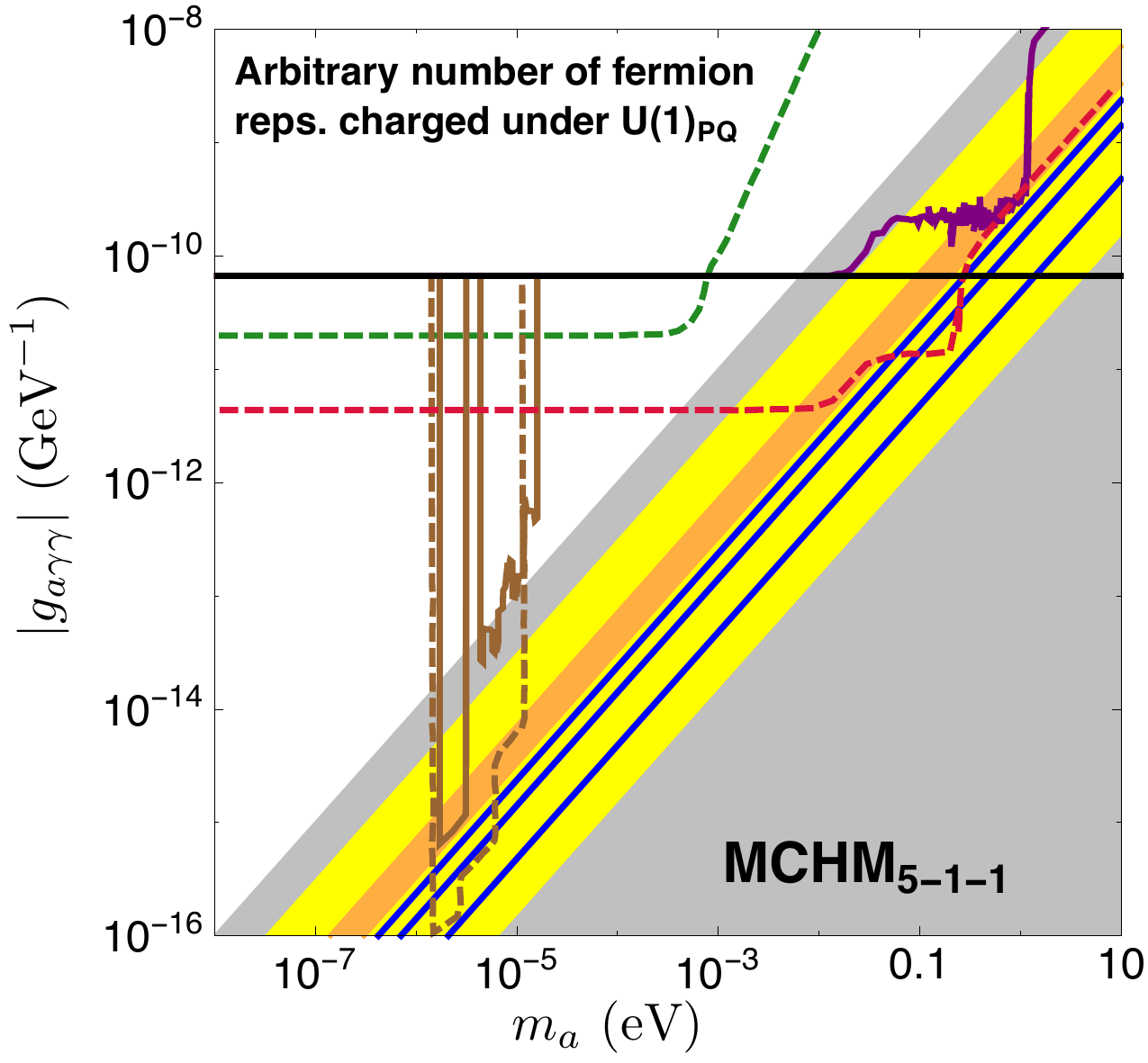}}
  \caption{Expected $g_{a\gamma\gamma}$ for KSVZ-type axionic extensions of the $SO(5) / SO(4)$ renormalizable Goldstone Higgs model described in Sec.~\ref{section_our_model}. The blue lines correspond to solutions in which the only PQ-chirally charged fermion(s) are $SO(5)$ singlet(s), e.g. Eq.~(\ref{natural-sol}), {\it and} which allow  $\mathcal{O}(f_a)$ fermion mass and  $\mathcal{O}(1)$ couplings: amongst the two uppermost lines, the upper (lower) one is the bottom (top) sector solution, while the extra one in Fig.~\ref{ma_ga_ParamSpace:sub2} corresponds to charging both sectors. Current limits from CAST~\cite{Anastassopoulos:2017ftl}, ADMX~\cite{Asztalos:2009yp, Carosi:2013rla, DePanfilis:1987dk, Wuensch:1989sa, Hagmann:1990tj} and and horizontal branch (HB) stars~\cite{Ayala:2014pea} are delimited by solid lines, while projected sensitivities for ALPS-II~\cite{Bahre:2013ywa}, IAXO~\cite{Carosi:2013rla ,  Irastorza:2011gs} and ADMX~\cite{Asztalos:2003px} are dashed. }
\end{figure}

In Fig.~\ref{ma_ga_ParamSpace:sub1} we project the values of $E/N$ obtained in Tab.~\ref{TableOnlyOneS} on  the $|g_{a\gamma\gamma}|$ versus $m_a$ parameter space  (see Eq.~\eqref{agammagamma_coupling}), depicting as a yellow band the region allowed when only one fermion representation of the composite Higgs model is allowed to be charged under PQ. 
This region is delimited by $E/N = (8/3, 76/15)$.
This is also the range if only the solutions with one exotic fermion chirally charged are taken into account,  as depicted by the orange hatched region superimposed. Would, instead, only solutions with one SM fermion chirally charged be considered, the region allowed would be smaller, corresponding to limiting values of $E/N$ $(8/3,2/3)$.
For comparison,  the grey band shows the expectations of the traditional KSVZ invisible axion models with only one exotic fermion charged under PQ, as updated recently in Ref.~\cite{DiLuzio:2016sbl} corresponding to values of $|g_{a\g\g}|$ delimited by $E/N=(5/3,44/3)$.     
   
The figure illustrates that, when only one fermion representation is charged under PQ, the region allowed by the renormalizable Goldstone Higgs model with minimal exotic fermion spectra {\it \`a la partial compositeness} discussed in this section is much narrower than that for KSVZ scenarios, a fact that should be relevant  for experimental searches. The reason is  that in the former models the charges of the exotic fermions are constrained via their essential participation in generating the light fermion masses, while in traditional KSVZ scenarios those charges are free, as light fermion masses result from the SM Yukawa couplings, which do not participate in the PQ mechanism. We will further deepen below on the underlying rationale, when letting all fermions  be arbitrarily charged under PQ.

 \FloatBarrier

\subsection{Arbitrary number of  fermions charged} 
   
Charging more than one  fermion expands logically the range of possible $E/N$ values. As an example,  $E/N=5/3$ when $\psi^{(1)}_L$, $\psi^{(1)\prime}_L$ , $\Lambda_3$ and $\Lambda_3'$ are charged under $U(1)_{\mathrm{PQ}}$ (and still $y_2=y_2'=0$ required by PQ invariance).   We consider next the general case in which  fields and couplings are allowed to take arbitrary PQ charges simultaneously (always as a function of just one field, the singlet $S$ or $S^\dagger$). The aim is to determine the maximum and minimum possible values of $|g_{a\g\g}|$.  Note that the condition $y_2=y_2'=0$ is no more necessary in the general case to obtain a PQ invariant setup involving the minimal set of exotic fermions responsible for light fermion masses, as Eq.~(\ref{Chiral_Differences}) can be fulfilled then even in the presence of only one scalar singlet.
For instance, with chirally charged exotic fermions it allows 
\begin{equation}\label{sinlabel}
\pm\beta(S)\equiv \beta(S_{M5})=-\beta(S_{M1}) \quad \Longrightarrow \Delta\psi^{(5)}=-\Delta\psi^{(1)}\,,
\end{equation}
suggesting a dynamical origin for both $M_1$ and $M_5$,\footnote{Charging  under PQ both $\psi^{(1)}$ and $\psi^{(5)}$ does not allow a natural solution with exotic fermion masses of $\mathcal{O}(f_a)$,  because of the constraints imposed by the top mass discussed earlier. The solution with small values for $\kappa_1$ and $\kappa_5$ is technically natural, though.} e.g. 
\begin{equation} \label{Replacement_with_y2}
M_5 \longrightarrow \kappa_5 \,S\,,\qquad M_1 \longrightarrow \kappa_1 \,S^\dagger\,.
\end{equation}
 The ensemble of solutions allowed by Eq.~(\ref{Chiral_Differences}) include  as well those in which {\it none} of the exotic fermions have PQ-chiral charges, that is, those in which the only fermions involved in the PQ mechanism are the SM ones. As previously stated, this interesting possibility exists for composite Higgs models while it is absent in KSVZ standard invisible axion models, and constitutes a distinctive feature.

Fig.~\ref{ma_ga_ParamSpace:sub2} depicts in yellow the generic band in parameter space allowed for arbitrary number of fermions chirally charged under PQ and arbitrary values of  $y_2$ and $y_2'$, whose limits correspond to $E/N=(2,56/3)$. A narrower orange band ($E/N=(11/3,17/3)$) 
 has been superimposed, in order to  indicate the smaller  parameter space of the solutions in which only exotic heavy fermions acquire  chiral PQ charges.       
For comparison, the grey region ($E/N<170/3$) is that for standard invisible axion models when they allow the simultaneous presence of several exotic fermions charged under the PQ symmetry, as recently predicted~\cite{DiLuzio:2016sbl}. This comparison reveals a striking fact: {\it while in the standard constructions it may be possible to make the strength of the axion-photon-photon coupling arbitrarily small, this is not possible in the wide range of Goldstone Higgs setups with fermionic partial compositeness reviewed here}.  
The generic origin of the narrower parameter space for composite Higgs models can be understood from Fig.~\ref{ma_ga_ParamSpace:sub2} as the net effect of two characteristics competing in opposite directions, 
      see Eqs.~(\ref{chain_renormalizable}) and (\ref{tbmass}):
      \begin{itemize}
 \item     Light fermion masses are directly mediated by the exotic fermions (while there are no SM Yukawa couplings), implying  strong constraints  on the possible PQ charges of the exotic dynamical mass parameters. They induce the very narrow orange band in Fig.~\ref{ma_ga_ParamSpace:sub2}. 
 \item The fact that SM fermions can now acquire PQ chiral charges (unlike in traditional KSVZ models) somewhat relaxes  the allowed parameter space. This explains the passage from the orange band to the wider yellow one in Fig.~\ref{ma_ga_ParamSpace:sub2}, in the most general case.   
      \end{itemize}
Overall, the comparison illustrates that, in the axion solutions of the renormalizable Goldstone Higgs models based on $SO(5)/SO(4)$ with minimal exotic fermion spectrum, the viable phenomenological parameter space is much restricted with respect to that for the standard invisible axion setups. We will see that this result holds as well for the many other Goldstone Higgs models in the literature to be discussed next.
  
\section{Extension by only one scalar 
     singlet: non-linear setups  } \label{section_models}

 In its strong coupling limit, the renormalizable   model discussed in the previous section (corresponding to a large mass for the SM scalar singlet $\sigma$ contained in its $SO(5)$ scalar five-plet) acquires a  non-linear formulation in terms of effective couplings, which is the usual approach for instance in  composite Higgs models.  In this non-linear context,  several very different fermionic UV contents have been considered in the literature.   This section will be entirely devoted to these effective non-renormalizable formulations. From the point of view of the effective field theory formulation, the implicit assumption is that the $\sigma$ particle of a putative renormalizable ultraviolet completion of composite Higgs models has been integrated out.
 
 The notation
MCHM$_{\rm{A-B-C}}$ is often used to indicate the fermionic spectrum
 of composite Higgs models, with $\rm{A, B, C}$ indicating the $SO(5)$ representation
 which contains the heavy partner of the SM doublet $q_L$, up-type right-handed and down-type
right-handed fermions, respectively.\footnote{Sometimes, when only one
subindex appears as in MCHM$_{\rm{A}}$ it is understood to be of the
type MCHM$_{\rm{A-A-A}}$.}  The heavy partner of a given SM quark is understood here as  
the SM multiplet contained in the $SO(5)$ exotic representation which is dominant in the generation of the SM quark mass, through a soft mixing $\Lambda_i$.
 For example, the model described in the previous section can be tagged in its fermionic content as MCHM$_{5-1-1}$ since the partners of the $q_L$ are found inside a five-plet of $SO(5)$ and those of $t_R$ and $b_R$ correspond to $SO(5)$ singlets; these partners were called $Q,\,Q',\,T$ and $B$ respectively and contained in $SO(5)$ five-plet and singlet representations, see  Eq.~\eqref{SU2_decomposition} and Tab.~\ref{Table_SMgauge_decomposition}.

We extend now the study performed in the previous sections to a plethora of fermionic spectra used usually in composite Higgs models~\cite{Carena:2014ria}, which are typically non-linear effective realizations. Details of the specific fermion representations involved are given in Tab.~\ref{Table_SMgauge_decomposition}, and the models 
 are  summarized in Tabs.~\ref{Table_model_content_1-5-10} and \ref{Table_model_content_14}.\footnote{Models
  with spinorial $SO(5)$ embeddings,
  e.g. MCHM$_4$~\cite{Agashe:2004rs}, are phenomenologically
  excluded in particular in view of $Z\to b
  \bar{b}$ data.\cite{Contino:2006qr}}
  For all models, the generation of the light quark masses results from a seesaw-like chain of interactions of the form 
 \beq \label{general_chain}
 q_L \xrightarrow[\Ll_q]{} Q_R \xrightarrow[M_Q]{} Q_L  \xrightarrow[y^t_1]{} T_L \xrightarrow[M_T]{} T_L \xrightarrow[\Ll_t]{} t_R\,,
 \eeq
where $\Lambda_i$ and $y_i$ generalize the MCHM$_{5-1-1}$ couplings in Eqs.~\eqref{SO5Lag} and \eqref{chain_renormalizable},  
upon the replacement  \{$ M_5 \rightarrow M_Q, M_1 \rightarrow M_T, \Ll_1 \rightarrow \Ll_q, \,\Ll_3 \rightarrow \Ll_t,\,\text{and}\,y_1\rightarrow y^t_1$\}.  An analogous chain holds for the bottom mass, with \{$ Q\rightarrow Q',\,T \rightarrow B,\,\Ll_q \rightarrow \Ll'_{q},\,\Ll_t\rightarrow \Ll_b\,, y_1^t\rightarrow y^b_1$\}.
The Yukawa-like couplings of exotic fermions to the Higgs particle, $y_t^1$ and $y_b^1$ (equivalent to $y_1$ and $y_1'$ in the notation used for the renormalizable model),  correspond to operators whose mass dimension is model dependent, as shown in Tabs.~\ref{Table_model_content_1-5-10} and \ref{Table_model_content_14}.   

We denote by $\Psi^{i}$ the $SO(5)$ representation which contains the heavy partner $i=Q,\,Q',\,T,\,B$, which in this study will be either a fermionic singlet, a five-plet, a ten-plet or a fourteen-plet, as shown in Tab.~\ref{Table_SMgauge_decomposition}.\footnote{The $Q,\,T,\,B$ representation superscript $(1)$, $(5)$ $(10)$ or $(14)$ shown in Tab.~\ref{Table_SMgauge_decomposition} are left implicit here for simplicity.} 
In MCHM$_{5-1-1}$ each of the four heavy partners ($Q,\,Q',\,T,\,B$) was contained in a different $SO(5)$ representation, so four exotic $SO(5)$ fermions were to be added, but this is not always needed as can be seen in Tab.~\ref{Table_SMgauge_decomposition}. For example,  MCHM$_{5-5-5}$ requires only two $SO(5)$ representations:  the $(5,2/3)$ representation $\psi^{(5)}$ contains both $Q$ and $T$, while the $(5,-1/3)$ representation $\psi^{(5)\prime}$ contains both $Q'$ and $B$. Indeed, in this model the SM-exotic fermion mixings are given by 
\bea
 \qbar_L Q_R \,&\supset\,   \qbar_L \spurion_{q} \Psi_R^{Q} \,&=\, \qbar_L \spurion_{2\times5} \psi_R^{(5)} \label{qmixing}\,, \\ 
\qbar_L Q'_R \,&\supset\,   \qbar_L \spurion_{q}' \Psi_R^{Q'} \,&=\,  \qbar_L \spurion_{2\times5} \psi_R^{(5)\prime} \,, \label{qprimemixing}\\ 
 \Tbar_L t_R \,&\supset\,   \Psibar^{T}_L \spurion_{t} t_R \,&=\, \psibar^{(5)}_L \spurion_{1\times5} t_R \label{tmixing}\,,  \\
 \Bbar_L b_R \,&\supset\,   \Psibar^{B}_L \spurion_{b} b_R\,&=\, \psibar^{(5)\prime}_L \spurion_{1\times5} b_R \,, \label{bmixing} 
\eea
 where again by $\spurion$ we denote dimensionless couplings, whose $SO(5)$ matrix dimension has been made explicit on 
the right-hand side for clarity. 
In summary,  $\Psi^{Q}=\Psi^{T}$ and $\Psi^{Q'}=\Psi^{B}$ in the MCHM$_{5-5-5}$ model. Yet other models shown  in Tabs.~\ref{Table_model_content_1-5-10} and \ref{Table_model_content_14} do not distinguish between $Q$ and $Q'$ and thus $\Psi^Q=\Psi^{Q'}$, further reducing the number of  exotic fermion representations required.

Generalizing the definitions above, the Lagrangian 
can be written as:
  \bea \label{models_lag}
  &&\mathcal{L}_{ferm} =\hspace{0cm}\bar{q}_{L} i\Ds  q_L + \bar{t}_R i\Ds  t_R + \bar{b}_R i\Ds  b_R  +\sum_{i=Q,\,Q',\,T,\,B} \Big\{ \bar{\Psi}^{i} i\Ds\Psi^{i} - \Big[\bar{\Psi}_L^{i} M_i\Psi^{i}_R +\hc\Big] \Big\} \\
 && -
  \L_{\text{Yuk.}} -\left[
 \Ll_{q} \,\qbar_L \,\spurion_q\, \Psi^{Q}_R \,+ \Ll'_{q}\, \qbar_L \,\spurion_{q}' \,\Psi^{Q'}_R \,+\, \Ll_{t}\, {\bar{\Psi}}^T_L \, \spurion_t\, t_R \,+ \,\Ll_{b}\, {\bar{\Psi}}^B_L\, \spurion_b \,b_R + \mathcal{L}_{\text{subdom.}}+ \hc \right]\,, \nn
 \eea
where the sum on the mass and kinetic terms runs over as many different fermion representations as needed, as discussed above. The dimensions of the $\spurion$ coupling matrices are model-dependent.

$ \L_{\text{Yuk.}}$ contains the low-energy effective fermion-Higgs operators of mass dimension $d\geq 4$ --this depends on the model-- which can be  schematically written as  
  \[
 \label{Lag-general}
\L_{\text{Yuk}} \sim \,\frac{y^t_1}{f^{n-1}} \Psibar^{Q}_L[\phi^n]  \Psi^{T}_R + \,\frac{y^b_1}{f^{n-1}} \Psibar^{Q'}_L [\phi^n] \Psi^{B}_R+\,\frac{y^t_2}{f^{n-1}} \Psibar^{Q}_R[\phi^n]  \Psi^{T}_L + \,\frac{y^b_2}{f^{n-1}} \Psibar^{Q'}_r [\phi^n] \Psi^{B}_L + \hc\,,
 \]
 where $\phi$ denotes here a five-component $SO(5)$ matrix with only four independent degrees of freedom, as its fifth component is fixed in the non-linear regime to be  $\sigma\to f^2 -2 |H|^2$, instead of  the dynamical field $\sigma$ of the previous renormalizable model Eq.~(\ref{SU2_decomposition}). The precise form of the $[\phi^n]$ insertions for each model considered can be read in Tabs.~\ref{Table_model_content_1-5-10} and \ref{Table_model_content_14} for illustration.   This Lagrangian generalizes  the Yukawa couplings of the renormalizable model in Eq.~(\ref{SO5Lag}), with the correspondence $\{y_1,\,y_1',\,y_2,\,y_2'\}\rightarrow \{y_1^t,\,y_1'^b,\,y_2^t,\,y_2'^b\}$.  Those models in Tabs.~\ref{Table_model_content_1-5-10} and \ref{Table_model_content_14} with Yukawa structures  of mass dimension four can be easily rewritten as renormalizable ones by simply replacing the non-linear constraint 
 mentioned above by a dynamical $\sigma$ field and adding a scalar potential, along the lines of the renormalizable model discussed in detail in Sect.~\ref{Model}; this is the case for instance of MCHM$_{5-10-10}$, MCHM$_{5-1-10}$ and MCHM$_{5-14-10}$, while an UV completion for models with higher-dimension Yukawa structures would require to consider extra mediator fields.  In any case, note that the precise form of the Yukawa structures is irrelevant for the $E/N$ values predicted, as the $SO(5)$ five-plet scalars $\phi$  are not PQ charged and in consequence that ratio only depends on the relative PQ chiral charges of fermions.\footnote{For some models involving 10-plets or 14-plets of $SO(5)$~\cite{Carena:2014ria}, the compact Lagrangian in Eq.~(\ref{Lag-general}) includes additional Yukawa structures with respect to those shown in Tabs.~\ref{Table_model_content_1-5-10} and \ref{Table_model_content_14}.  They do not make a difference for the $E/N$ values predicted.}
 
The last line in Eq.~(\ref{models_lag}) contains the mixings between SM and exotic fermions. Its first four terms are those participating in the chain in Eq.~\eqref{general_chain}, which gives the dominant contributions to the light fermion masses (the different content and $SO(5)$ matrix size of the $\spurion$ couplings are model-dependent and have been left implicit here for notational simplicity). 
 $\L_{\text{subdom.}}$ includes other fermion mixing terms which give subdominant contributions to the light fermion masses;\footnote{These were not made explicit in the summary of models in Ref.~\cite{Carena:2014ria} which focused on the issue of mass, but here their presence/absence does influence the size of the axion-$\gamma\gamma$ parameter space and we thus include them.} they are the equivalent of the $\Lambda_2$ and $\Lambda'_2$ couplings in the renormalizable model Eq.~(\ref{SO5Lag}) discussed in Sec.~\ref{section_our_model}. They are couplings of the type $\qbar_L  Q_R$, $\bar{T}_L  t_R$ or $\bar{B}_L b_R$. As an illustration, model $\text{MCHM}_{5-10-10}$ allows subdominant contributions of the form $\qbar_L \psi_R^{(10)}$ and $\bar{\psi}_L^{(5)}  t_R$ in addition to the dominant mixings $\qbar_L  \psi_R^{(5)}$, $\bar{\psi}_L^{(10)}  t_R$ and  $\bar{\psi}_L^{(10)}  b_R$, see Tabs.~\ref{Table_model_content_1-5-10} and \ref{Table_model_content_14}; using the $\Psi$ notation, they read
 \begin{equation}
\L_{\text{subdom.}}^{5-10-10}=\tilde{\Lambda}_{q} \qbar_L \spurion\, \Psi^T_R + \tilde{\Lambda}_t \Psibar^Q_L \spurion\, t_R\,.
\end{equation}
We have identified and shown in Tabs.~\ref{Table_model_content_1-5-10} and \ref{Table_model_content_14} the set of subdominant $\tilde{\Lambda}_i$  terms for each  of the models considered. These terms further constrain significantly the phenomenological axion-photon  analysis below. 

A $U(1)_{\mathrm{PQ}}$-invariant formulation of the 
effective Lagrangian in Eq.~(\ref{models_lag}) can be achieved along the same lines as for the renormalizable model in Sec.~\ref{section_our_model}. Scalar fields singlet under both $SO(5)$ and the SM gauge group and  whose vev sets the size of the PQ scale $f_a$ as in  Eq.~(\ref{Sdefinition}) are introduced, combined with the promotion to dynamical fields of some of the mass parameters described above, i.e. 
 \beq 
 \label{dynamic-nonlinear}
 M_i \rightarrow \kappa_i S_{Mi}\,,\qquad \text{and / or}\qquad \Ll_i \rightarrow \lambda_i S_{\Ll i}\,.
 \eeq
Again, the small $\kappa_i$ and $\lambda_i$ values, which may be required in order to get a spectrum of exotic fermion masses in the TeV range, are protected by $U(1)$ chiral symmetries under which only the fermions transform. In some cases, $\mathcal{O}(1)$ parameters may be safely allowed as previously discussed in Sec.~\ref{section_our_model}.~\footnote{As this case corresponds to singlet fermion masses much higher than the $\Lambda_s$ scale, the effective Lagrangian formulation in Eq.~(\ref{Lag-general}) should have to be replaced then by one in which the heavy singlet fermion fields are not present. Their effect will be included in higher dimension operators resulting from the integration of those fermions. For the practical analysis here there is no need of expliciting these steps. Additionally, the caveats discussed in the introduction and after Eq.~(\ref{promoting2}) as to the stability of the $f$ scale for these solutions are also pertinent here.} Alike to Eq.~\eqref{Chiral_Differences}, 
 the PQ chiral charge differences  are then given by 
 \bea
 \label{SM_chiral_differences}
&\,&\Delta \Psi^i = \beta(S_{Mi})\,,\nn \\
 &\,&\Delta t = \beta(S_{\Lambda q}) + \beta(S_{\Ll t}) - \beta(S_{MQ}) -\beta(S_{MT})\,,
 \eea
and analogously for the bottom sector.  In the minimal extension scenario of enlarging the spectrum by only one singlet scalar $S$, 
the Yukawa couplings may force some of the  scalar PQ charges $\beta(S_{Mi})$  to vanish, see Tabs.~\ref{Table_model_content_1-5-10} and \ref{Table_model_content_14}.   

A  clarification is pertinent from the point of view of the effective field theory. Although Eq.~(\ref{dynamic-nonlinear}) is written in terms of a  scalar singlet under the SM and $SO(5)$, this is only for bookkeeping and easy comparison with the renormalizable model in the previous section. The full $S$ dynamics is not playing a role in the phenomenological analysis, or the maybe more complex UV completion for that matter. The only ingredient used is the promotion of dimensional parameters to dynamical ones, endowing them with PQ charges as in Eq.~(\ref{SM_chiral_differences}), and the only field retained is the light axion stemming from them. In other words, the analysis is independent of the physics of the real components of $S$.
 
 \vspace{0.5cm}
\begin{table}[h!] \small
\begin{center}
\begin{tabular}{  |>{$}c<{$} | >{$}c<{$} |>{$}c<{$} | >{$}c<{$}  | >{$}c<{$} |  >{$}c<{$} |   } 
\hline
\text{MCHM}& \begin{matrix}\Psi^Q \\ \Psi^{Q' } \end{matrix}& \begin{matrix}\Psi^T \\  \Psi^B  \end{matrix} & \mathcal{L}_{Yuk.} &\L_{\text{subdominant}}& \Big[\frac{E}{N}\Big|_{|g_{a\gamma\gamma}| \,\text{min.}}, \frac{E}{N}\Big|_{|g_{a\gamma\gamma}| \,\text{max.}} \Big]   \\ [2mm]
\hline
\hline
\multirow{2}{*}{$5-1-1$} & (5,\,2/3)& (1,\,2/3)& \psibar_L^{(5)} \phi \,\psi_R^{(1)}&\psibar_L^{(5)}(\spurion_{5\times 1} t_R )&  {\multirow{2}{*}{[2,\,56/3]}}   \\[2mm]
				& (5, \,-1/3) & (1,\,-1/3) & \psibar_L^{'(5)} \phi \,\psi_R^{'(1)} &\psibar_L^{'(5)}(\spurion_{5\times 1}b_R)& \\ [2mm]
\hline 
\multirow{2}{*}{$5-5-5$ } & (5,\,2/3)& (5,\,2/3)& \psibar_L^{(5)} \phi \phi^\dagger \psi_R^{(5)}&\multirow{2}{*}{$-$} & \multirow{2}{*}{[2,\,{-4}/{3}]}  \\[2mm]
				& (5, \,-1/3) & (5,\,-1/3) & \psibar_L^{'(5)} \phi \phi^\dagger \psi_R^{'(5)} && \\ [2mm]
\hline 
\multirow{2}{*}{$5-10-10$ } &\multirow{2}{*}{(5,\,2/3)}& \multirow{2}{*}{(10,\,2/3)}& \multirow{2}{*}{$\psibar_L^{(5)} \phi\,  \psi^{(10)}_R$}&(\qbar_L \spurion_{2\times 10}) \psi_R^{(10)}   &  \multirow{2}{*}{[2,\,{50}/{3}]}   \\[2mm]				&  					&  					&& \psibar_L^{(5)}(\spurion_{5\times 1} t_R ) & \\
				\hline 
\multirow{2}{*}{$10-10-10$ } &\multirow{2}{*}{(10,\,2/3)}& \multirow{2}{*}{(10,\,2/3)}& \multirow{2}{*}{$\phi^\dagger \bar{\psi}^{(10)}_L   \psi^{(10)}_R \phi$}&\multirow{2}{*}{$-$}  & \multirow{2}{*}{[2,\,{2}/{3}]}   \\[2mm]
				&  					&  					&&  & \\
				\hline 
\multirow{2}{*}{$10-5-10$ } & \multirow{2}{*}{(10,\,2/3)}& (5,\,2/3)& \bar{\psi}^{(10)}_L \phi  \,\psi_R^{(5)}& (\qbar_L \spurion_{2\times 5}) \psi_R^{(5)}  &  \multirow{2}{*}{[2,\,{2}/{3}]}   \\[2mm]				& 								 & (10,\,2/3) & \phi^\dagger \bar{\psi}^{(10)}_L   \psi^{(10)}_R \phi & \psibar_L^{(10)}(\spurion_{10\times 1} t_R )   &\\ [2mm]
\hline 
\multirow{2}{*}{$5-5-10$ } & \multirow{2}{*}{(5,\,2/3)}& (5,\,2/3)& \psibar_L^{(5)} \phi \phi^\dagger \psi_R^{(5)}&(\qbar_L \spurion_{2\times 10}) \psi_R^{(10)}  &  \multirow{2}{*}{[2,\,{2}/{3}]}   \\[2mm]
				& 						 & (10,\,2/3) &  \psibar_L^{(5)} \phi  \, \psi^{(10)}_R &\psibar_L^{(10)}(\spurion_{10\times 1} t_R )  &\\ [2mm]
\hline 
\multirow{3}{*}{$5-1-10$ } & \multirow{3}{*}{(5,\,2/3)}& (1,\,2/3)& \psibar_L^{(5)} \phi \,\psi_R^{(1)}&\multirow{3}{*}{$\begin{matrix} (\qbar_L \spurion_{2\times 10}) \psi_R^{(10)}  \\  \psibar_L^{(5)}(\spurion_{5\times 1} t_R )  \\ \psibar_L^{(10)}(\spurion_{10\times 1} t_R )\end{matrix}$} & \multirow{3}{*}{[2,\,12]}   \\[2mm]
				& 						 & (10,\,2/3) &  \psibar_L^{(5)} \phi \,  \psi^{(10)}_R  & &\\ [2mm]
				& 						 & 		&  & &\\ [0mm]
\hline  
\end{tabular}
\end{center}
\caption{Summary of the non-renormalizable MCHMs in the literature which involve only fermionic five-plets and/or ten-plets and/or singlets. The second and third columns specify the particle content of each model; the fourth contains the $SO(5)-$invariant Yukawa interactions (the first row inside each column is proportional to $y^t_1$ and the second to $y^b_1$, except when they coincide). The fifth column specifies subdominant mixing terms. The last column gives the ranges of $E/N$ that define the phenomenological band in the $|g_{a\g\g}|$ versus $m_a$ parameter space.}\label{Table_model_content_1-5-10}
\end{table}

\vspace{0.5cm}
\begin{table}[h!] \small
\begin{center}
\begin{tabular}{  |>{$}c<{$} | >{$}c<{$} |>{$}c<{$} | >{$}c<{$}  | >{$}c<{$} |  >{$}c<{$} |   } 
\hline
\text{MCHM}& \begin{matrix}\Psi^Q \\ \Psi^{Q' } \end{matrix}& \begin{matrix}\Psi^T \\  \Psi^B  \end{matrix} & \mathcal{L}_{Yuk.} &\L_{\text{subdominant}}& \Big[\frac{E}{N}\Big|_{|g_{a\gamma\gamma}| \,\text{min.}}, \frac{E}{N}\Big|_{|g_{a\gamma\gamma}| \,\text{max.}} \Big]   \\ [2mm]

\hline
\hline
\multirow{3}{*}{$14-1-10$ } & \multirow{3}{*}{(14,\,2/3)}& (1,\,2/3)	&   \phi^\dagger \psibar_L^{(14)}   \psi_R^{(1)} \phi	&\multirow{3}{*}{$\begin{matrix} (\qbar_L \spurion_{2\times 10}) \psi_R^{(10)}  \\  \psibar_L^{(10)}(\spurion_{10\times 1} t_R )  \\ \psibar_L^{(14)}(\spurion_{14\times 1} t_R ) \end{matrix}$}										&     \multirow{3}{*}{[2,\,{158}/{3}]}   \\[2mm]
				& 						 & (10,\,2/3) 	&   \phi^\dagger \psibar_L^{(14)} \psi_R^{(10)} \phi 	&			&  \\ [5mm]
\hline
\multirow{4}{*}{$14-5-10$} & \multirow{4}{*}{(14,\,2/3)}		&\multirow{2}{*}{(5,\,2/3)}	& \multirow{2}{*}{$\psibar_L^{(14)} \phi \,\psi_R^{(5)}$ }				& (\qbar_L \spurion_{2\times 5}) \psi_R^{(5)}	&  {\multirow{4}{*}{[2,\,-100/3]}}   \\[2mm]
					& 	 					&  						& 									 					& (\qbar_L \spurion_{2\times 10}) \psi_R^{(10)}	& \\ [2mm]
					& 						& \multirow{2}{*}{(10,\,2/3)} 	&	\multirow{2}{*}{$\phi^\dagger \psibar_L^{(14)}  \,\psi_R^{(10)} \phi$}	&\psibar_L^{(14)}(\spurion_{14\times 1}t_R)	& \\ [2mm]
					& 						& 						& 														&\psibar_L^{(10)}(\spurion_{10\times 1}t_R)	& \\ [2mm]
\hline 
\multirow{4}{*}{$5-14-10$} & \multirow{4}{*}{(5,\,2/3)}		&\multirow{2}{*}{(14,\,2/3)}		& \multirow{2}{*}{$\psibar_L^{(5)} \phi \,\psi_R^{(14)}$ }	& (\qbar_L \spurion_{2\times 14}) \psi_R^{(14)}	&  {\multirow{4}{*}{[2,\,29/3]}}   \\[2mm]
					& 	 					&  						& 									 		& (\qbar_L \spurion_{2\times 10}) \psi_R^{(10)}	& \\ [2mm]
					& 						& \multirow{2}{*}{(10,\,2/3)} 	&\multirow{2}{*}{$\psibar_L^{(5)} \phi \,\psi_R^{(10)}$}	&\psibar_L^{(5)}(\spurion_{5\times 1}t_R)	& \\ [2mm]
					& 						& 						& 											&\psibar_L^{(10)}(\spurion_{10\times 1}t_R)	& \\ [2mm]
\hline 
\multirow{2}{*}{$10-14-10$ } 	& \multirow{2}{*}{(10,\,2/3)}	& (14,\,2/3)		&\phi^\dagger \bar{\psi}^{(10)}_L   \psi^{(14)}_R \phi		& (\qbar_L \spurion_{2\times 14}) \psi_R^{(14)}   	&  \multirow{2}{*}{[2,\,2/3]}   \\[2mm]	
						& 						& (10,\,2/3) 		& \phi^\dagger \bar{\psi}^{(10)}_L   \psi^{(10)}_R \,\phi 	& \psibar_L^{(10)}(\spurion_{10\times 1} t_R )    	&\\ [2mm]
\hline
\multirow{2}{*}{$14-10-10$ } 	& \multirow{2}{*}{(14,\,2/3)}	& \multirow{2}{*}{$(10,\,2/3)$}	&  \multirow{2}{*}{$\phi^\dagger \bar{\psi}^{(14)}_L  \,\psi_R^{(10)} \phi $}	& (\qbar_L \spurion_{2\times 10}) \psi_R^{(10)}  &  \multirow{2}{*}{[2,\,83/3]}   \\[2mm]		
						& 						& 					 	& 														& \psibar_L^{(14)}(\spurion_{14\times 1} t_R )   &\\ [2mm]
\hline 
\multirow{2}{*}{$14-14-10$ } 	& \multirow{2}{*}{(14,\,2/3)}	& (14,\,2/3)	&  \phi^\dagger \bar{\psi}^{(14)}_L  \,\psi_R^{(14)} \phi 	& (\qbar_L \spurion_{2\times 10}) \psi_R^{(10)}  &  \multirow{2}{*}{[2,\,2/3]}   \\[2mm]		
						& 						& (10,\,2/3) 	&  \phi^\dagger \bar{\psi}^{(14)}_L  \,\psi_R^{(10)} \phi 	& \psibar_L^{(10)}(\spurion_{10\times 1} t_R )   &\\ [2mm]
\hline 
\end{tabular}
\end{center}
\caption{Summary of the non-renormalizable MCHMs in the literature that include fermions in the 14-plet representation of SO(5). The second and third columns specify the particle content of each model; the fourth contains the $SO(5)-$invariant Yukawa interactions (the first row inside each column is proportional to $y^t_1$ and the second to $y^b_1$, except when they coincide). The fifth column specifies subdominant mixing terms. The last column gives the ranges of $E/N$ that define the phenomenological band in the $|g_{a\g\g}|$ versus $m_a$ parameter space.}\label{Table_model_content_14}
\end{table}

The ensuing general expression for the ratio of electromagnetic and color anomalies $E/N$ reads now 
\begin{equation}
\frac{E}{N}
=\frac{2}{3}\frac{182\,\Delta\psi^{(14)}+94 \,\Delta\psi^{(10)}+38 \Delta\psi^{(5)}+23\Delta\psi^{'(5)}+4\Delta\psi^{(1)}+\Delta\psi'^{(1)}+4\Delta t + \Delta b}{14 \Delta\psi^{(14)}+10 \Delta\psi^{(10)}+5 \Delta\psi^{(5)}+5\Delta\psi^{'(5)}+\Delta\psi^{(1)}+\Delta\psi^{'(1)}+\Delta t+\Delta b}\,.
\end{equation}
which generalizes Eq.~(\ref{general_E/N}) derived for the  MCHM$_{5-1-1}$ model.

\begin{figure}[t]
\centering
  \includegraphics[width=0.85\linewidth]{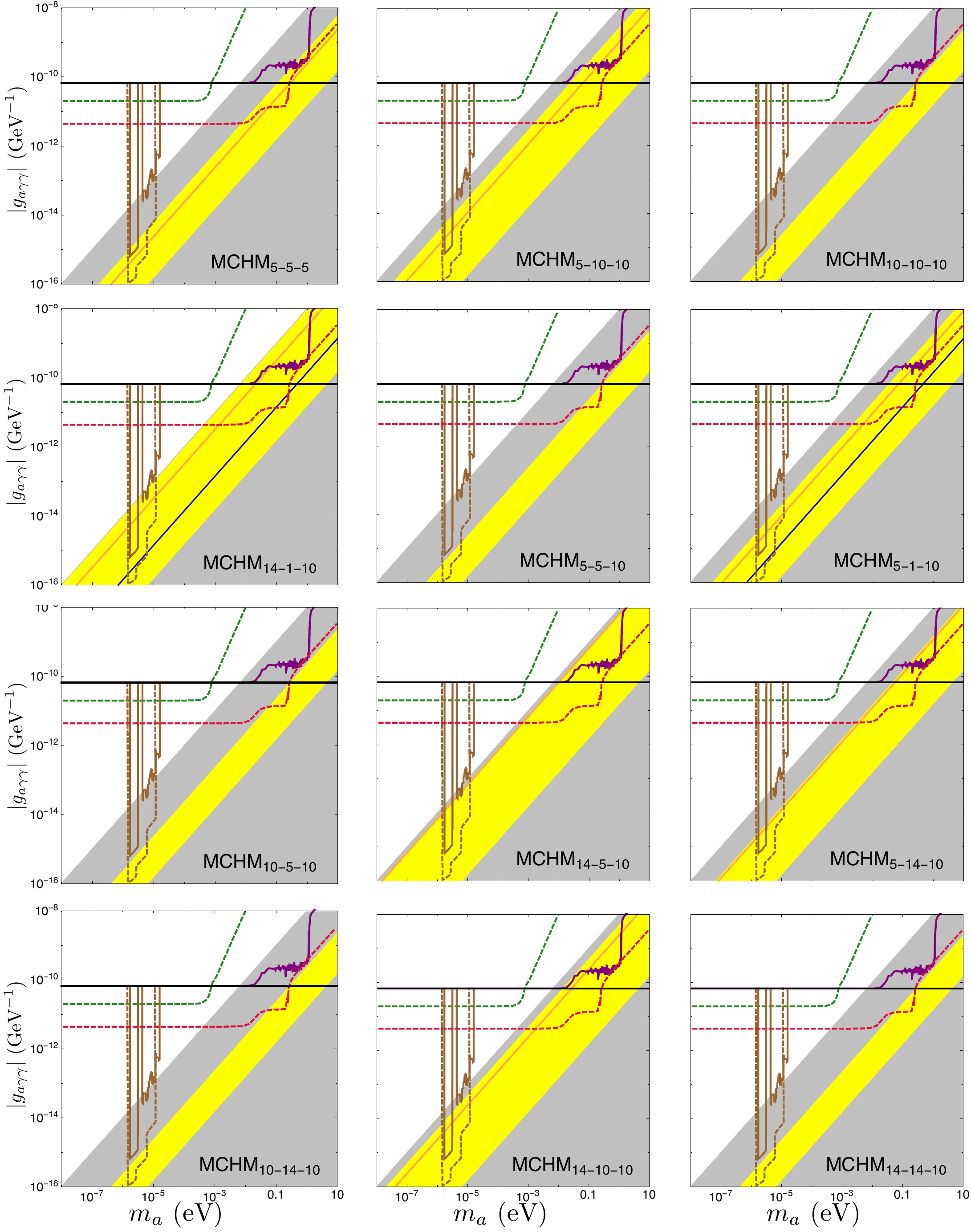}
  \includegraphics[width=0.75\linewidth]{CompleteLegend.pdf}
  \caption{Yellow bands: expected $g_{a\gamma\gamma}$ for various MCHMs  when extending the spectrum by only one singlet scalar field PQ charged. 
  Both dominant and subdominant heavy-SM fermion mixings are included, see  Eq.~\eqref{models_lag}. The subset of solutions in which only exotic fermions are charged  is depicted by orange bands/lines. The blue lines correspond to solutions with only  $SO(5)$-singlet fermions charged, for which masses $\mathcal{O}(f_a)$ and  couplings of $\mathcal{O}(1)$ could be allowed. For comparison, the predictions of the updated~\cite{DiLuzio:2016sbl} standard KSVZ invisible axion scenarios are depicted as a grey region. The plot for the case MCHM$_{5-1-1}$ is not shown here because it is exactly the same as that shown in Fig.~\ref{ma_ga_ParamSpace:sub2}.}
\label{Bands_many_models}
\end{figure}

Using the results above, we have identified the $E/N$ values that  correspond to the maximum and minimum possible  values of $|g_{a\gamma\gamma}|$, for the different minimal (in fermion content) models in Tabs.~\ref{Table_model_content_1-5-10} and \ref{Table_model_content_14}, within the minimal extension of the spectrum by just one scalar singlet and allowing all fermions to take arbitrary PQ charges. The results 
are shown on the last column of the table, and 
the corresponding allowed area of the $(m_a,\,g_{a\gamma\gamma})$ plane  is depicted for the different models by yellow bands in  Figs.~\ref{ma_ga_ParamSpace:sub2} and \ref{Bands_many_models}.\footnote{The results are independent of the linear or non-linear formulation of the models, assuming that no scalar acquires is PQ-charged other than the added singlet $S$ .}  

The allowed yellow regions  tend to be wider for the models which involve a  number of different exotic fermion representations, as otherwise the constraints implied by their Yukawa couplings reduce strongly the parameter space of PQ-invariant formulations. The extreme case is that in which only one exotic  $SO(5)$ representation is involved, as the Yukawa coupling forces then its chiral PQ charge to vanish (alike to the constraint imposed  in traditional KSVZ theories by the SM Yukawa couplings) and the remaining allowed parameter space is  entirely due to  PQ-charged SM fermions. Again, the overall pattern is that the parameter space corresponding to PQ chirally charged exotic fermions is strongly constrained, while the presence of PQ chirally charged SM fermions relaxes  that constraint to some extent.
A narrower orange band has been superimposed over the yellow ones in Figs.~\ref{ma_ga_ParamSpace:sub2} and \ref{Bands_many_models} for illustration, indicating the smaller parameter space that would remain if only the exotic fermions   would be allowed to acquire chiral PQ charges:  the figure shows that MCHM$_{5-5-5}$, MCHM$_{10-10-10}$, MCHM$_{10-5-10}$, MCHM$_{5-5-10}$, MCHM$_{10-14-10}$ and MCHM$_{14-14-10}$ would not be then compatible with a minimal PQ invariant formulation. There is no good reason for such restriction to non PQ chirally charged SM fermions within composite Higgs models, though, so those models are also good candidates for an axion solution in the framework of a pGB nature for the Higgs boson.

Figs.~\ref{ma_ga_ParamSpace:sub2} and \ref{Bands_many_models}  also depict in grey the area allowed by the recent updated predictions of the traditional KSVZ invisible axion model. Overall, the comparison 
shows that the phenomenological region allowed by general Goldstone Higgs realizations with minimal fermion content {\it \`a la partial compositeness} is much more restrictive, and thus predictive, than for traditional KSVZ constructions, confirming  the pattern already identified for the renormalizable model in the previous section.

Finally, the very few particular cases with $\mathcal{O}(1)$  $\kappa_i$ or $\lambda_i$ parameters are indicated in  Figs.~\ref{ma_ga_ParamSpace:sub2} and \ref{Bands_many_models} by blue lines superimposed over the bulk of the solutions. Only MCHM$_{5-1-1}$,  MCHM$_{5-1-10}$ and MCHM$_{14-1-10}$ allow this possibility, being the only ones containing at least one heavy partner in a singlet representation of $SO(5)$. This is needed for the mixings between light and heavy fermions to be exclusively $SO(5)$-invariant, e.g. $\psi^{(1)}t_R$ or $\psi^{(1)}b_R$,  allowing the singlet $\psi^{(1)}$ to be charged under PQ and its mass term promoted to a dynamical field, without promoting to scalar fields any of the soft-breaking couplings. It could be a natural possibility that the  fermionic fields which are singlets of $SO(5)$ acquire a mass much larger than that of the $SO(5)$ group whenever a new higher physics scale is present, as for instance $f_a$ in the framework of a $U(1)_{\mathrm{PQ}}$ solution to the strong CP problem.

\vspace{0.3cm}
 The model which overall allows for a larger variety of implementations is MCHM$_{5-1-1}$, see Fig.~\ref{ma_ga_ParamSpace:sub2}, because it has the largest number of different fermionic representations, which translates into a sizable fraction of models with only exotic fermions charged (orange band) and three solutions with $\mathcal{O}(f_a)$ exotic fermion masses.

\FloatBarrier

\section{Summary and Outlook}
\label{Sect:Conclusions}

An important problem of current dynamical solutions to the strong CP problem, assuming only the SM gauge symmetries, is that they are strongly fine-tuned, as the axion scale $f_a$ is phenomenologically required to be many orders of magnitude above the electroweak scale, while the scalar sector of the models communicates both scales and tends to homogenize their values. This  problem hinders all invisible axion constructions.

With this perspective, we have explored the implementation of the Peccei-Quinn axial symmetry  $U(1)_{\mathrm{PQ}}$  in models in which the Higgs particle has a pseudo-Nambu-Goldstone boson nature. In them,  that Higgs ancestry results from some global symmetry spontaneously broken at high-energy, protecting the Higgs mass from electroweak hierarchy issues, as it can only become massive after some small and explicit symmetry breaking. Furthermore, 
 the  global symmetry forbids  direct SM Yukawa couplings. The light observed fermion masses are then generated via ``partial compositeness'': a seesaw-like pattern mediated by heavy exotic fermion partners of the SM fermions; Yukawa couplings are allowed by the global symmetry only for the partners. In general, these exotic fermions appear in vectorial representations of the SM gauge group. This means that, by construction, Goldstone Higgs models come with a heavy spectrum alike to that of the hadronic invisible axion model KSVZ. We have discussed possible extensions of their spectra so as to make those models  $U(1)_{\mathrm{PQ}}$ invariant, with the minimality criterion of {\it} not extending their fermionic sector, and focusing  on the simplest case of  $SO(5)$ global symmetry. 
 
{\it We have shown that the minimal extension consisting in the addition of a single $SO(5)$ scalar singlet to the spectrum and no additional heavy fermions suffices to implement the PQ symmetry in those models}, although the constraints for extensions with more than one singlet have also been determined. In a first step, a renormalizable sigma model with MCHM$_{5-1-1}$ fermionic content has been thoroughly explored, which allowed a precise identification of model building constraints.  From the point of view of naturalness, the Peccei-Quinn scale $f_a$ may be expected to be close to the mass of the sigma particle (as neither is protected by the symmetries of the problem and the scalar potential may connect them), which when taken very massive results in the customary low-energy effective  non-linear formulation typical of effective Goldstone Higgs constructions.  We recall that, as in the QCD linear and non-linear $\sigma$, a heavy $\sigma$ can be obtained without destabilising the Higgs mass nor the EW scale~\cite{Feruglio:2016zvt}.
 In a second step, a plethora of fermionic setups used in non-renormalizable formulations existing in the literature has been considered. The latter differ by the type of exotic heavy fermions and Yukawa couplings,  and we have discussed how to formulate them as renormalizable sigma models and how to extend them minimally (only by scalar singlets) so as to acquire a Peccei-Quinn invariant formulation. 
 
The issue of naturalness for the solutions found has been discussed in detail and used as a discriminating tool. Although the Higgs mass is protected from the electroweak hierarchy problem by construction, the question is pertinent with respect to the other scales of the theories, given the large value of $f_a$. 
When all heavy exotic fermion mass eigenstates are assumed to remain at most of the order of the composite scale ($\sim 1-100$ TeV), we have found that all axion solutions are technically natural as they are protected by a chiral symmetry under which some fermions transform but not the scalars. An appealing and different possibility has been also identified, though, for a very small subset of the solutions found: that in which the $SO(5)$ singlet fermion representations --often used in the literature in Goldstone Higgs models-- may be the only ones with Peccei-Quinn charges and having masses of order $f_a$. Only three of the many fermionic setups considered satisfy this more restrictive criteria: MCHM$_{5-1-1}$,  MCHM$_{5-1-10}$ and MCHM$_{14-1-10}$, and for each of them some solution(s) could accommodate very heavy fermions.
Such options could not  necessarily require small dimensionless parameters among the axionic couplings discussed here; this would be a natural solution in the sense that the mass parameters for $SO(5)$ singlets are not protected by any low-energy symmetry. Although the value of the Higgs mass itself cannot be destabilized by large contributions due to the heavy $SO(5)$ singlet fermions, it remains to be clarified, though, whether such heavy singlet fields may destabilize instead the value of the $f$ scale or not, and whether a fine tuning of other parameters in the model (e.g. in the scalar potential) would be required to compensate for this effect.

The phenomenological predictions for the axion-photon-photon coupling (actively searched for at present by many experiments all over the world) have been next determined for all models explored. We have demonstrated that {\it the region in the $(m_a,\,g_{a\gamma\gamma})$ parameter space allowed for Goldstone Higgs models on which the PQ symmetry is implemented without enlarging their original fermionic sector is much more restrictive than that for standard invisible axion formulations}. The reason is that, while in the latter scenarios the SM fermion masses are unrelated to the vectorial exotic sector, in Goldstone Higgs scenarios the generation of SM masses via partial compositeness imposes stringent relations  among the parameters and couplings of the exotic fermions that mediate them.
For instance, within the fermionic spectra of existing Goldstone Higgs models, that is assuming as fermionic content exclusively their inherent minimal spectrum, it is not possible to obtain an arbitrarily small axion-photon-photon coupling for any given $m_a$. The latter would require instead   
to add extra fermions to the spectrum with the specific purpose of implementing the axion solution via the couplings of those extra fermions. This restricted parameter space for the minimal fermionic setup holds in spite of the  complex  spectrum of fermionic spectra in Goldstone Higgs models, which in general requires several distinct fermion representations, in contrast to recent finds for standard invisible axion models~\cite{DiLuzio:2016sbl, DiLuzio:2017pfr}.
  
It is remarkable that the plethora of existing Goldstone Higgs models exhibit by construction a KSVZ-like structure simply with their inherent minimal fermionic sector, a suggestive fact explored here.
Although the precise phenomenological analysis has been done for the case of only adding a single scalar singlet field \, the underlying reason for the restricted parameter space is generic and should hold with more extended scalar spectra. 
This  enhanced predictivity of the minimal Goldstone Higgs setups explored has a relevant impact on the planned experimental searches,  and may also serve as discriminating tool in case of future axion and/or Goldstone Higgs signals.
 \vspace{-.2cm}
\section*{Acknowledgments}
\vspace{-.2cm}
We acknowledge Mary K. Gaillard, Luca di Luzio, Pedro Machado, Luca Merlo, Pablo Qu\'ilez, Stefano Rigolin and Ver\'onica Sanz for very interesting conversations and comments. 
M.B.G and R.dR acknowledge Berkeley LBNL, where part of this work has been developed, for hospitality during their visit in October 2017.
This project has received funding from the European Union's Horizon 2020 research and innovation programme under the Marie Sklodowska-Curie grant agreements No 690575  (RISE InvisiblesPlus) and  No 674896 (ITN ELUSIVES). M.B.G., R.dR, S. Pascoli and S. Saa also acknowledge support from the 
 the Spanish Research Agency (Agencia Estatal de Investigaci\'on) through the grant IFT Centro de Excelencia Severo Ochoa SEV-2016-0597. 
M.B.G., R.dR and S. Saa   acknowledge financial support from the "Spanish Agencia Estatal de Investigaci\'on" (AEI) and the EU ``Fondo Europeo de Desarrollo Regional'' (FEDER) through the project FPA2016-78645-P. 
I.B. acknowledges support from the Villum Foundation, NBIA, the Discovery Centre at Copenhagen University and the Danish National Research Foundation (DNRF91)
S.P.  would like to acknowledge partial
support from the European Research Council under ERC Grant ``NuMass'' (FP7-
IDEAS-ERC ERC-CG 617143), and from the Wolfson Foundation and the Royal Society and also acknowledges  IFT-UAM/CSIC  and SISSA for support and hospitality during part of this work. The work of S.S. was supported through the grant BES-2013-066480 of the Spanish MICINN.


\providecommand{\href}[2]{#2}\begingroup\raggedright\endgroup

\end{document}